\documentclass[fleqn,usenatbib]{mnras}

\usepackage[T1]{fontenc}
\usepackage{ae,aecompl}

\usepackage{graphicx}	
\usepackage{amsmath}	
\usepackage{amssymb}	
\usepackage[export]{adjustbox}
\usepackage{pdflscape}
\usepackage{float}
\usepackage{subfig}

\title[Mapping the ISM in SDP.81]{Full of Orions: a 200-pc mapping of the interstellar medium in the redshift-3 lensed dusty star-forming galaxy SDP.81}

\author[M. Rybak et al.]{
Matus Rybak$^{1}$\thanks{E-mail: mrybak@strw.leidenuniv.nl},
J. A. Hodge$^{1}$,
S. Vegetti$^{2}$,
P. van der Werf$^{1}$,
P. Andreani$^{3}$,
\newauthor
L. Graziani,$^{4,5}$
and J. P. McKean$^{6}$
\\
$^{1}$ Leiden Observatory, Leiden University, Niels Bohrweg 2, 2333 CA Leiden, the Netherlands\\
$^{2}$ MPA Garching, Karl-Schwarzschild Stra{\ss}e 2, 85748 Garching bei M{\"u}nchen, Germany\\
$^{3}$ European Southern Observatory, Karl-Schwarzschild Stra{\ss}e 2, 85748 Garching bei M{\"u}nchen, Germany\\
$^{4}$ Dipartimento di Fisica, Sapienza, Universit$\grave{a}$ di Roma, Piazzale Aldo Moro 5, 00185, Roma, Italy\\
$^{5}$ INAF/Osservatorio Astrofisico di Arcetri, Largo E. Femi 5, 50125 Firenze, Italy\\
$^{6}$ Kapteyn Astronomical Institute, University of Groningen, Postbus 800, NL-9700 AV Groningen, the Netherlands\\
$^{7}$ ASTRON, Netherlands Institute for Radio Astronomy, Oude Hoogeveensedijk 4, 7991 PD Dwingeloo, the Netherlands
}

\date{
Accepted 2020 March 17. Received 2020 March 5; in original form 2019 September 30.}

\pubyear{2020}

\begin{document}
\label{firstpage}
\pagerange{\pageref{firstpage}--\pageref{lastpage}}
\maketitle

\begin{abstract}
We present a sub-kpc resolved study of the interstellar medium properties in SDP.81, a $z=3.042$ strongly gravitationally lensed dusty star-forming galaxy, based on high-resolution, multi-band ALMA observations of the FIR continuum, CO ladder and the [\ion{C}{ii}] line. Using a visibility-plane lens modelling code, we achieve a median source-plane resolution of $\sim$200~pc. We use photon-dominated region (PDR) models to infer the physical conditions -- far-UV field strength, density, and PDR surface temperature -- of the star-forming gas on 200-pc scales, finding a FUV field strength of $\sim10^{3}-10^{4} G_0$, gas density of $\sim10^5$ cm$^{-3}$ and cloud surface temperatures up to 1500~K, similar to those in the Orion Trapezium region. The [\ion{C}{ii}] emission is significantly more extended than that FIR continuum: $\sim$50~per cent of [\ion{C}{ii}] emission arises outside the FIR-bright region. The resolved [\ion{C}{ii}]/FIR ratio varies by almost 2~dex across the source, down to $\sim2\times10^{-4}$ in the star-forming clumps. The observed [\ion{C}{ii}]/FIR deficit trend is consistent with thermal saturation of the C$^+$ fine-structure level occupancy at high gas temperatures. We make the source-plane reconstructions of all emission lines and continuum data publicly available.
\end{abstract}




\section{Introduction}

Dusty star-forming galaxies (DSFGs) with star-formation rates (SFR) of 100-1000~M$_\odot$ yr$^{-1}$ play a key role in the epoch of peak star-forming activity of the Universe, in the $z=2-4$ redshift range (e.g., \citealt*{Casey2014, Swinbank2014}). The rest-frame far-UV radiation from the young, massive stars is often completely obscured by their large dust reservoirs and re-radiated at rest-frame far-infrared (FIR) wavelengths. 

The key to understanding the intense star-formation in DSFGs is their interstellar medium (ISM) - dense gas and dust in their (giant) molecular clouds (GMCs). In the canonical GMC picture (e.g., \citealt{Tielens1985}), the surface of the clouds is exposed to the intense far-UV radiation from the new-born massive O/B-type stars. This results in a thermal and chemical stratification of the GMC: while deep in the cloud, the gas is in the molecular form (H$_2$, CO); the ionizing flux and gas temperature increase towards the surface, causing CO to dissociate into C and O, with C being ionized into C$^+$ in the photon-dominated region (PDRs) at the cloud surface. Crucially, as the depth of individual layers -- and the strength of the emission lines associated with individual species -- is largely determined by the surface the FUV field strength ($G$) and the cloud density ($n$(H)), observing emission from multiple chemical species (e.g., C$^+$, C and O fine-structure and CO rotational transitions, dust continuum) allows the PDR properties to be inferred, typically via forward modelling.

The PDR properties in present-day (U)LIRGS have been extensively studied using FIR and mm-wave spectroscopy. An early study by \citet{Stacey1991} -- combining \textit{Kuiper Airborne Observatory} [\ion{C}{ii}] and FIR continuum observations with ground-based CO(1--0) observations -- found a range of densities $n$(H)$=10^3-10^6$~cm$^{-3}$, with $G$ in excess of $10^3$~$G_0$\footnote{$G_0$ denotes the Habing field, $1\,G_0 = 1.6\times10^{-3}$ erg s$^{-1}$ cm$^{-2}$.}. These studies were revolutionized by \textit{Herschel} FIR spectroscopy, opening doors to large systematic surveys of FIR emission lines down to $\sim$500-pc scales in nearby star-forming galaxies and (ultra)luminous infrared galaxies ((U)LIRGS), e.g., \citet{GraciaCarpio2011, Diaz2013, Rosenberg2015, Smith2017, Diaz2017, Herrera2018a}. 

Previous studies of gas and dust properties in DSFGs have been limited to spatially-integrated quantities, due to their faintness ($S_\mathrm{850~\mu m}$ of few mJy) and compact sizes (few arcsec), compared to the resolution available at FIR/mm-wavelengths. \citet{Stacey2010} used unresolved [\ion{C}{ii}], CO(2--1)/(1--0) and FIR continuum observations of a sample of twelve $z=1-2$ galaxies to infer typical $G=10^{2.5}-10^{4}~G_0$, $n$(H)$\simeq10^3-10^{4.5}$~cm$^{-3}$. More recently, the large samples of strongly-lensed DSFGs discovered in FIR/mm-wave surveys by \textit{Herschel} - H-ATLAS \citep{Eales2010, Negrello2010, Negrello2017} and HerMES \citep{Oliver2012, Wardlow2012} - and the South Pole Telescope \citep{Vieira2013, spilker2016} have provided an opportunity to obtain robust continuum and line detections in a fraction of time compared to non-lensed sources. For example, \citet{Gullberg2015} used ground-based FIR, [\ion{C}{ii}] and low-$J$ CO emission to infer the ISM properties in a sample of strongly lensed galaxies from the South Pole Telescope sample. \citet{Brisbin2015} used \textit{Herschel} [\ion{O}{i}] 63-$\mu$m and ground-based [\ion{C}{ii}] and low-$J$ CO observations to infer $n$(H)$=10^3-10^2$~cm$^{-3}$, $G=10^1-10^3~G_0$ in a sample of eight $z$=1.5-2.0 DSFGs. \citet{Wardlow2017} used \textit{Herschel} fine-structure line spectroscopy of strongly lensed DSFGs, obtaining $n$(H)$=10^1-10^3$~cm$^{-3}$, $G=10^{2.2}-10^{4.5}~G_0$. Similarly, \citet{Zhang2018b} used \textit{Herschel} [\ion{N}{ii}] and [\ion{O}{i}] observations for the H-ATLAS DSFGs to infer typical densities of $n$(H)$=10^3-10^4$~cm$^{-3}$. 

The main limitation of deriving PDR properties from spatially-integrated observations is the implicit assumption that the PDR conditions are uniform across the source, i.e., that the FIR continuum and different emission lines are co-spatial. However, resolved observations of DSFGs found that low-$J$ CO emission is significantly more extended than the FIR continuum (e.g., \citealt{Ivison2011, Riechers2011, Spilker2015, CR18}), and that mid-/high-$J$ CO emission is more compact than low-$J$ CO emission (e.g., \citealt{Ivison2011}, R15b). Additionally, line ratios in strongly lensed sources can be affected by differential magnification, introducing a significant bias in the inferred source properties, particularly for highly-magnified galaxies (e.g., \citealt{Serjeant2012}). 

Consequently, resolved multi-tracer observations are necessary to robustly infer the PDR conditions in DSFGs. For example, \citet{Rybak2019} have used resolved [\ion{C}{ii}], CO(3--2) and FIR continuum ALMA observations to study the properties of the star-forming ISM in the central regions ($R\leq2$~kpc) of two $z\sim3$ DSFGs from the ALESS sample \citep{Hodge2013,Karim2013}, finding FUV fields in excess of $10^4$~G$_0$ and moderately high densities $n$(H)=$10^{3.5}-10^{4.5}$~cm$^{-3}$. In this regime, the surface temperature of the PDR regions is of the order of few hundred K and the [\ion{C}{ii}] emission becomes thermally saturated, resulting in a pronounced [\ion{C}{ii}]/FIR deficit \citep{MunozOh2016}.

In this work, we use high-resolution, multi-Band ALMA observations to map the ISM properties in the strongly lensed $z=3.042$ DSFG SDP.81 at 200-pc resolution to address the following questions:
\begin{itemize}
\item How are different tracers of ISM - FIR continuum, CO and [\ion{C}{ii}] emission spatially distributed?
\item What is the cooling budget of the molecular gas, and does it vary across the source?
\item What are the conditions of the star-forming gas in DSFGs at $z\sim3$ - in particular, FUV field strength, density and surface temperature of the PDR regions? Do they vary significantly across the source?
\end{itemize}

This paper is structured as follows: in Section~\ref{sec:observations}, we summarize the ALMA observations analyzed in this work; Section~\ref{sec:imaging+modelling} describes the synthesis imaging and lens-modelling of this data. In Section~\ref{sec:results+discussion}, we use resolved FIR continuum, CO and [\ion{C}{ii}] maps to investigate the dust properties of SDP.81 (Section~\ref{subsec:dust_sed}), CO spectral energy distribution and the gas cooling budget (Section 4.2) and the [\ion{C}{ii}]/FIR ratio and deficit (Section 4.3). Section~5 then describes the PDR modelling procedure and results, and discusses the resolved ISM conditions in SDP.81 and the physical mechanisms driving the [\ion{C}{ii}]/FIR deficit. Finally, Section~\ref{sec:conclusions} summarizes the results of this work.

Throughout this paper we use a flat $\Lambda$CDM cosmology from \citet{Planck2015}. Adopting the spectroscopic redshift of $z=3.042$ \citep{alma2015}, this translates to a luminosity distance of 26480~Mpc, and an angular scale of 1~arcsec = 7.857~kpc \citep{Wright2006}.

\section{Observations}
\label{sec:observations}

\subsection{Target description}
\label{subsec:target}
SDP.81 (J2000 09:03:11.6 +00:39:06; \citealt{Bussmann2013}), is a $z_S=3.042$ DSFG, strongly lensed by a foreground early-type galaxy with $z_L=0.299$. Identified in the H-ATLAS survey \citep{Eales2010, Negrello2010, Negrello2017}, SDP.81 was targeted by the ALMA 2014 Long Baseline Campaign \citep{alma2015}. Ancillary observations include HST near infra-red imaging \citep{Dye2014}, \textit{Herschel} FIR spectroscopy \citep{Valtchanov2011, Zhang2018b}, and radio/mm-wave low-resolution spectroscopy and imaging \citep{Valtchanov2011, Lupu2012}. This extensive high-resolution, high-S/N ancillary data make SDP.81 perhaps the best-studied DSFG.

\citet{Rybak2015a,Rybak2015b} (henceforth R15a, R15b) presented the reconstructed maps of Bands~6 and 7 dust continuum, and CO(5--4) and (8--7) emission with a typical resolution of 50-100~pc, alongside the reconstruction of \textit{Hubble Space Telescope} near-IR imaging. These revealed a significantly disturbed morphology of stars, dust and gas in SDP.81, with an ordered rotation in a central dusty disk some 2--3~kpc; a similar conclusion has been reached by independent analysis both in the $uv$-plane \citep{Hezaveh2016} and {\sc Clean}ed images \citep{Dye2015, Swinbank2015}. Based on the kinematic analysis of the CO(5--4) and (8--7) maps, SDP.81 is classified as a post-coalescence merger (\citealt{Dye2015}, R15b, \citealt{Swinbank2015}). In this paper, we adopt the CO(5--4) and CO(8--7) from R15b, and obtain matched-resolutions reconstructions of Band~6 and Band~7 continuum (Section~3.1.3).

\subsection{ALMA Band~3 observations}
The CO(3--2) line was observed in ALMA Cycle 5, (Project \#2016.1.00633). The observations were carried out on 2017 September 9 in ALMA Band~3. While the original project requested a total of 4~hours on-source with a resolution of 0.2~arcsec, the total observations amount to only $\sim$45~min. The array configuration consisted of 40 12-meter antennas, with baselines extending from 39 to 7550~m, covering angular scales between $\sim$0.10 and 19~arcsec. The measured precipitable water vapour was 1.5~mm. The frequency setup consisted of four spectral windows (SPWs), centered at 97.730, 99.626, 85.813 and 87.614 GHz, respectively. Two SPWs were configured in the line mode with 1920 channels of 976.6~kHz width per channels, for a total bandwidth of 1.875~GHz per SPW. The other two SPWs were configured in the continuum mode (channel width 15.625~MHz), with a bandwidth of 2.0~GHz per SPW.

\subsection{ALMA Band~8 observations}

The [\ion{C}{ii}] line was observed in ALMA Cycle~4 (Project \#2016.1.01093.S). The observations were carried out on 2016 November 14 with 44 12-meter antennas with baselines extending from 13 to 850~m, corresponding to angular scales between $\sim$0.16 and 10.5~arcsec. The precipitable water vapour ranged between 0.52--0.64~mm. The frequency setup consisted of four SPWs, centered on 468.35, 458.24, 456.35 and 470.24~GHz, respectively. SPWs 0--2 were configured in a continuum mode with a bandwidth of 2.0~GHz per SPW, while SPW~3 was split into 240 7.812-kHz-wide channels, for a total bandwidth of 1.875~GHz.

\subsection{ALMA 2014 Long Baseline Campaign observations}
\label{subsec:ALMA_LBC}
As the ALMA Long Baseline Campaign (LBC) observations of SDP.81 were described in detail in \citet{alma2015}, here we provide a only brief overview of the data The observations comprised of 5.9, 4.4 and 5.6 hours on-source in Bands~4, 6 and 7 respectively, taken in October 2014. The frequency setup covered the CO(5--4) line in Band 4, CO(8--7) and H$_2$O (2$_{02}-1_{11}$) lines in Band 6 and CO(10--9) line in Band 7. The frequency setup provided 7.5~GHz bandwidth per Band, with a channel width of 0.976 -- 1.95~MHz for line-containing spectral windows (SPWs), and 15.6~MHz for the continuum SPWs. The antenna configuration consisted of 22-36 12-meter antennas, with baselines ranging from $\sim$15~m to $\sim$15~km, resulting in a synthesized beam size of $\sim30$~mas for Band 7 to $\sim$55~mas for Band 4 (Briggs weighing, robust = 1). For the CO lines, a strong drop in S/N was observed for baselines longer than 1~M$\lambda$ \citep{alma2015}. This resulted in restricting the CO line analysis to baselines $\leq1$~M$\lambda$; here we adopt this $uv$-distance cut for all the ALMA datasets.

\section{Data processing and analysis}
\label{sec:imaging+modelling}

\subsection{Data preparation and imaging}

All the data processing was done with {\sc Casa} versions 4.7 and 5.0 \citep{McMullin2007}. Figure~\ref{fig:cleaned} shows the {\sc Clean}ed frequency-integrated maps of the Band 8, [\ion{C}{ii}], CO(3--2) and CO(10--9) emission, obtained using natural weighting and an outer Gaussian taper of 0.2~arcsec.

\subsubsection{ALMA Band~3}
The data were calibrated and flagged using the standard ALMA Cycle~4 pipeline; in addition, antennas DA53 and DV22 were manually flagged due to a systematically high antenna temperature. 
The CO(3--2) line was detected between 85.509 and 85.571~GHz. Due to the low S/N, we frequency-averaged the line-containing channels into a single frequency bin. A time-averaging of 20~s was applied, as in R15a,b. This reduces the size of the problem by a factor of a few without compromising the surface-brightness sensitivity. The {\sc Clean}ed images (Figure~\ref{fig:cleaned}) show CO(3--2) emission along the main arc and in the counterimage, with a peak S/N of 5.6$\sigma$. 

The {\sc Clean}ed Band 3 continuum map has an rms noise of 10~$\mu$Jy beam$^{-1}$ (natural weighting). We do not detect any Band~3 continuum emission from the lensed source at 3$\sigma$ significance. We detect the active galactic nucleus (AGN) synchrotron emission in the lensing galaxy (rest-frame frequency 111~GHz) with a total flux of 52$\pm$10~$\mu$Jy. The observed flux is consistent with the \citet{Tamura2015} synchrotron spectrum for the foreground AGN within $2\sigma$.

\subsubsection{ALMA Band~8}
\label{subsec:imaging_band8}
After applying the standard ALMA Cycle~4 calibration, we performed manual flagging in the time-domain to remove several integrations which showed elevated noise levels; a total of 6 minutes of data was flagged. We also flagged channels affected by atmospheric lines at 468.0 and 457.3~GHz.

The [\ion{C}{ii}] line is detected between 469.791 and 470.253~GHz. The observed image-plane [\ion{C}{ii}] integrated line flux of 170$\pm$20 Jy km~s$^{1}$ ($(2.7\pm0.3)\times10^{-18}$ W~m$^{-2}$) is consistent with the \textit{Herschel} Spire FTS measurement of $(2.9\pm0.6)\times10^{-18}$~W m$^{-2}$ \citep{Zhang2018b} within 1$\sigma$. This confirms that ALMA Band~8 observations are not resolving out extended [\ion{C}{ii}] emission\footnote{Note that an older reduction of the \textit{Herschel} [\ion{C}{ii}] spectra by \citet{Valtchanov2011} yielded a 5$\times$ higher an integrated [\ion{C}{ii}] line flux.}.

The [\ion{C}{ii}] absorption from diffuse ionized gas was detected towards some Galactic star-forming regions (e.g., \citealt{Gerin2015}). At high redshift, \citet{Nesvadba2016} detected a [\ion{C}{ii}] absorption towards to $z=3.4$ DSFG, which they interpreted as evidence for infalling gas. Looking for a potential [\ion{C}{ii}] absorption in the {\sc Clean}ed cube with a velocity resolution of 40~km s$^{-1}$, we do not find any evidence for [\ion{C}{ii}] absorption against the 160-$\mu$m continuum.

\subsubsection{ALMA Long Baseline Campaign: Bands 6 and 7 continuum, CO(10--9) line}

To ensure that all the tracers are compared on the same spatial scales, we revisit the \citet{alma2015} Bands 6 and 7 continuum, applying a $uv$-distance cut of $\leq$1~M$\lambda$ as opposed to the 2-M$\lambda$ cut in R15a. For details of the continuum data preparation, we refer the reader to R15a.

\begin{figure*}
\begin{centering}
\includegraphics[width=12 cm, clip=true]{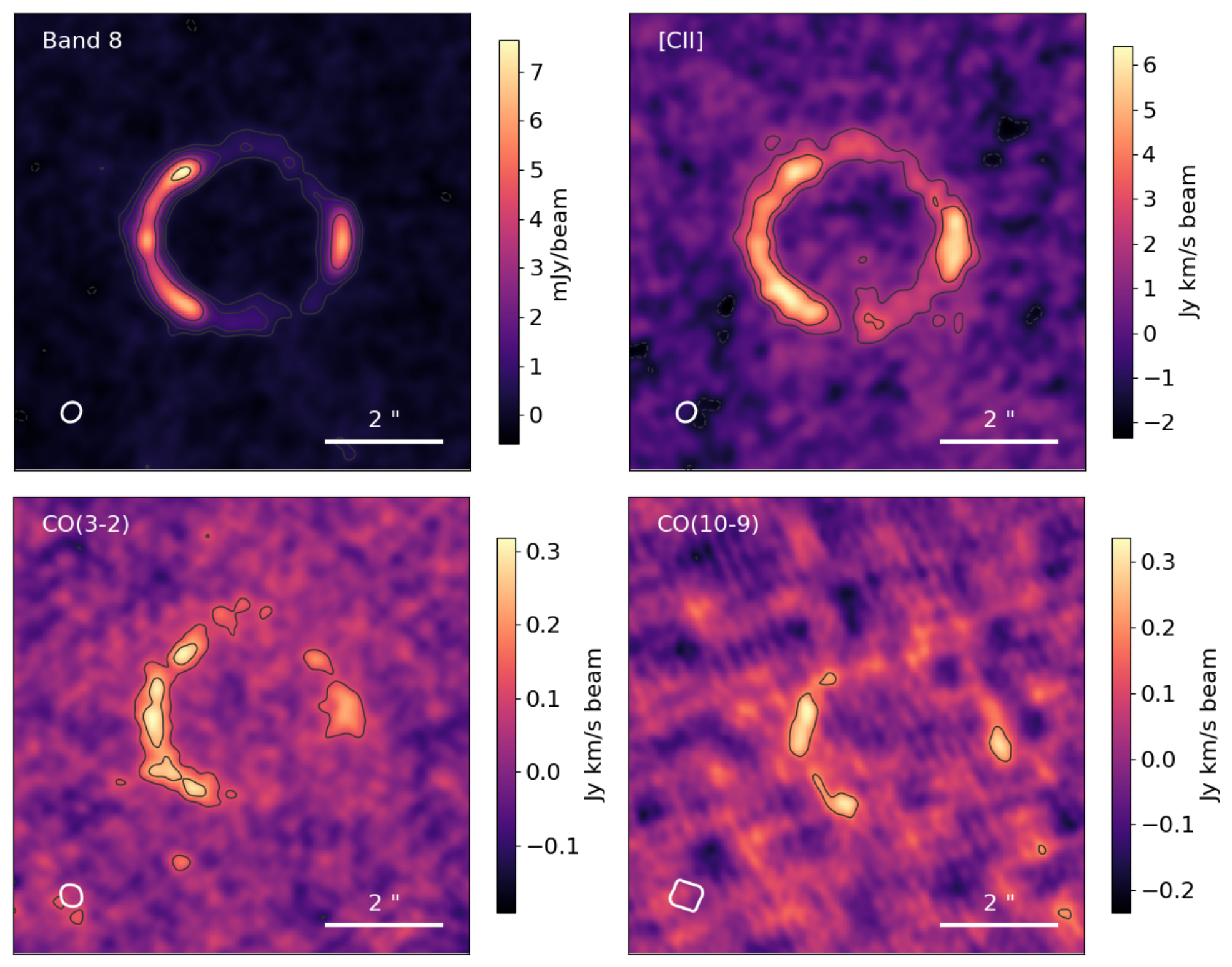}

\caption{ALMA imaging of SDP.81: Band~8 (rest-frame 160~$\mu$m) continuum, [\ion{C}{ii}], CO(3--2) and CO(10--9) lines, all based on naturally-weighted data. The contours start at $\pm3\sigma_\mathrm{rms}$ level, increase in steps of $3\sigma_\mathrm{rms}$ and are truncated at 18$\sigma_\mathrm{rms}$. The synthesized beam FWHM is indicated by the ellipse in the lower left corner. \label{fig:cleaned}}
\end{centering}
\end{figure*}

Finally, we include the CO(10--9) line, which was observed in ALMA Band~7 (see \citealt{alma2015} for detailed description and imaging). The CO(10--9) line is detected between 284.733--285.52~GHz; due to the low S/N ratio (peak image-plane SN$\simeq5$, Fig.~\ref{fig:cleaned}) we frequency-averaged the visibilities into a single channel. Although we obtain an integrated CO(10--9) line flux (measured from the {\sc Cleaned} images) $\sim$50~per cent higher than reported by \citet{alma2015} (derived from spectral fitting), the two measurements are consistent within 2$\sigma$.

\begin{table*}
\caption{Observed [\ion{C}{ii}], CO(3--2) and (10--9) line and Band~8 flux densities, measured from the naturally-weighted, velocity-averaged {\sc Clean}ed images. Individual columns list the observed central frequency $\nu_\mathrm{obs}$, synthesized beam FWHM size and position angle (east of north), rms noise of the {\sc Clean}ed images $\sigma_\mathrm{rms}$, total flux-density $S_\nu$ and maximum surface brightness $S_\mathrm{peak}$, line width at zero intensity $\Delta v$, and line flux $S_\nu \Delta v$ (if applicable). \label{tab:image-plane}}
\begin{center}
 \begin{tabular}{@{}l|cccccccc @{}}
 \hline \hline
Line & $\nu_\mathrm{obs}$ & Beam FWHM & Beam PA & $\sigma_\mathrm{rms}$ & $S_\nu$ & $S_\mathrm{peak}$ & $\Delta v$ & $S_\nu \Delta v$ \\
 & [GHz] & [arcsec] & [deg] & [mJy beam$^{-1}$] & [mJy] & [mJy beam$^{-1}$] & [km s$^{-1}$] &[Jy km s$^{-1}$]\\ 
 \hline
$[$\ion{C}{ii}$]$ & 470.02 & 0.32$\times$0.26 & 138 & 0.5 & 180$\pm$7 & 5.5 & 700$\pm$100 & 170$\pm$20\\
CO(3-2) & 85.55 & 0.36$\times$0.24 & 137 & 0.13 & 17.9$\pm$1.8 & 0.73 & 660$\pm$110 & 11.8$\pm$2.3\\
CO(10--9) & 285.1 & 0.31$\times$0.19 & 24 & 0.07 & 6.7$\pm$0.9 & 0.33 & 780$\pm$100 & 4.5$\pm$0.6\\
Band 8 cont. & 463.35 & 0.32$\times$0.26 & 138 & 0.14 & 112$\pm$2.0 & 6.8 & -- & --\\
 \hline
 \end{tabular}
\end{center}
\end{table*}

\subsection{Lens modelling}
\label{subsec:lens_modelling}
We reconstruct the source-plane surface brightness distribution of individual tracers via lens-modelling, thereby minimizing the differential magnification bias. The additional complication in case of interferometric data arises from the fact that radio-interferometers do not measure directly the sky-plane surface brightness distributions, but the visibility function, which is essentially a sparsely-sampled Fourier transform of the sky. Furthermore, the high angular resolution provided by ALMA requires the source to be reconstructed on a pixellated grid, as simple analytic models for the source (e.g., S\'ersic profiles) are no longer sufficient to fully capture the complex structure of the source. Consequently, several visibility-fitting lens-modelling codes aimed primarily at ALMA observations, have been developed (R15a, \citealt{Hezaveh2016, Dye2018}).

We perform the lens-modelling using the method first presented in \citet{Rybak2015a}, which is an extension of the Bayesian technique of \citet{Vegetti2009} to the radio-interferometric data. To summarize, this technique uses a parametric lens model, while the source is reconstructed on an adaptive triangular grid obtained via the Delaunay tessellation. This setup provides an increased source-plane resolution in the high-magnification regions. We minimize the following penalty function, which combines $\chi^2$ calculated in the visibility-space with a source-plane regularization to constrain the problem and prevent noise-fitting:

 \begin{multline} 
P(\boldsymbol{s}\,|\, \psi(\boldsymbol{\eta},\boldsymbol{x}), \lambda_s, \boldsymbol{d})=(\mathbf{FL} \boldsymbol{s} - \boldsymbol{d})^{\top} \mathbf{C}_d^{-1} (\mathbf{FL}\boldsymbol{s} - \boldsymbol{d})+\\
+\lambda_s (\mathbf{R}_s \boldsymbol{s})^{\top} \mathbf{R}_s \boldsymbol{s},
\label{equ:penalty_f}
 \end{multline}
 
where $\mathbf{F}$ and $\mathbf{L}$ are Fourier-transform (including the primary beam response) and lensing operators, $\lambda_s$ and $\mathbf{R}_s$ are the regularisation constant and the regularisation matrix for the source surface brightness distribution and $\mathbf{C}_d^{-1}$ is the covariance matrix of the observed visibilities. For a detailed discussion of the lens-modelling technique, we refer the reader to \citet{Koopmans2005} and \citet{Vegetti2009}.

We adopt the lens model of R15a derived from 0.1-arcsec resolution ALMA Band~6 and 7 continuum observations, which follows a singular isothermal ellipsoid profile with an external shear component. Keeping the lens mass model fixed, we re-optimize for the offset between the lens center and phase-tracking centre (which differ for different observations) and the source regularization parameter $\lambda_S$. The sky-plane pixel size is set to 25~mas; as the adaptive grid is obtained by casting back every second pixel, the synthesized beam is sub-sampled by a factor of $\sim4$. The primary beam response is approximated by an elliptical Gaussian. The rms noise for each visibility is estimated as the $1\sigma$ scatter of the real/imaginary visibilities for a given baseline over a 10-minute interval, rather than from the {\sc Sigma} column of the ALMA Measurement Set files.

We estimate the uncertainty on the reconstructed source-plane surface brightness distribution by drawing 1000 random realizations of the noise map given the $\sigma_\mathrm{rms}$ for each baseline, solving for the source using the best-fitting lens model and $\lambda_S$, and calculating the scatter per grid element from the resulting source reconstructions.

As the source-plane surface brightness sensitivity depends on the spatial position, we estimate the spatially-integrated magnifications for each tracer as follows: we take all the source-plane pixels with S/N $\geq$1, 2 and 3, forward-lens them into the image-plane and calculate the corresponding magnification and uncertainty for each S/N threshold; and finally estimate the overall magnification as the mean of the magnification factors for different S/N thresholds.

\subsection{Source-plane reconstructions}

We now discuss the source-plane reconstructions for the ALMA Band~8 continuum and the [\ion{C}{ii}] and CO(3--2) emission. Given the varying S/N of individual line and continuum observations, we further assess the reliability of the source-plane reconstruction by performing our lens modelling analysis on mock datasets in Appendix~\ref{sec:appendix_B}; these confirm the robustness of our source-plane reconstruction, with the mean surface brightness recovered within 10--15~per cent for most datasets ($\sim$20~per cent for CO(3--2)), comparable to the ALMA flux calibration uncertainty. We therefore consider our reconstructions to be robust.

\begin{figure*}

\includegraphics[width=18 cm, clip=true, left]{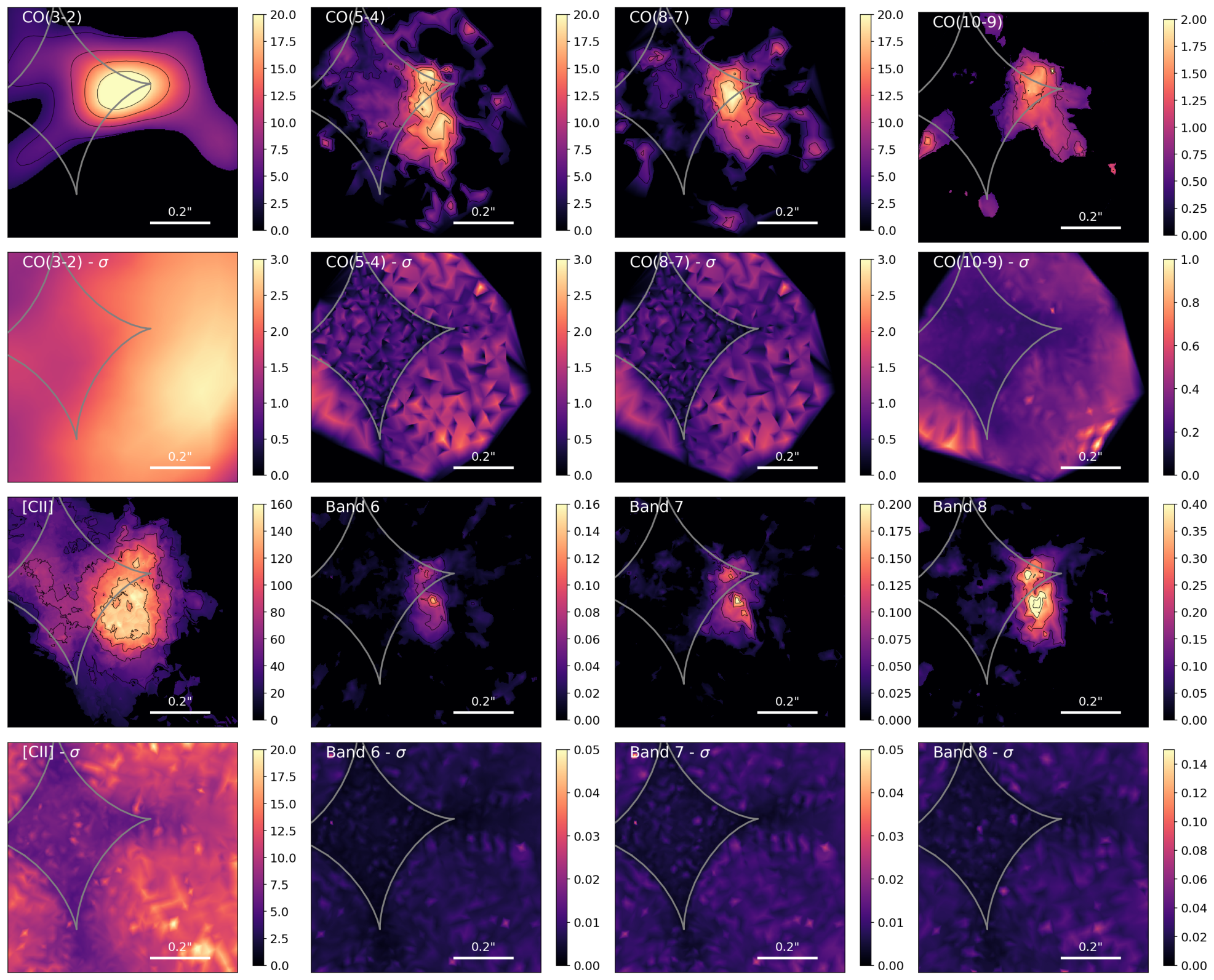}
\caption{Source-plane reconstructions of [\ion{C}{ii}], CO(3-2), CO(5-4), (8-7), and (10-9) line emission (units mJy km~s$^{-1}$~kpc$^{-2}$), and ALMA Bands 6, 7 and 8 continuum (units mJy kpc$^{-2}$), with associated $1\sigma$ uncertainty maps; the CO(5--4) and CO(8--7) data are adopted from R15b. In the source reconstruction maps, we masked regions with S/N$\leq$3. Grey lines indicate the lensing caustics. The scattered emission on the edges of the field is a modelling artifact. Note the large extent of the CO and [\ion{C}{ii}] emission compared to the FIR-bright source. The source-plane reconstructions and the associated errors are available online. \label{fig:reconstructions}}
\end{figure*}

Fig.~\ref{fig:reconstructions} presents the best-fitting models of the [\ion{C}{ii}], 160-$\mu$m continuum and CO(3--2) emission; the full output of the lens modelling procedure are provided in Appendix~\ref{sec:appendix_A}. The source-plane models reproduce the full range of image-plane structures (given the S/N), as demonstrated by the lack of significant residuals in the dirty-image plane. The median source-plane resolution is $\sim$30~mas, which corresponds to $\sim$200~parsec; the resolution in the vicinity of the caustic is as high as $\sim$3~pc. To allow the comparison on a uniform spacial scale, for the analysis in Section~\ref{sec:results+discussion} we re-sample the reconstructed source onto a Cartesian grid with 200$\times$200~pc pixels.

\begin{figure*}
\begin{centering}
\includegraphics[width=14cm, clip=true]{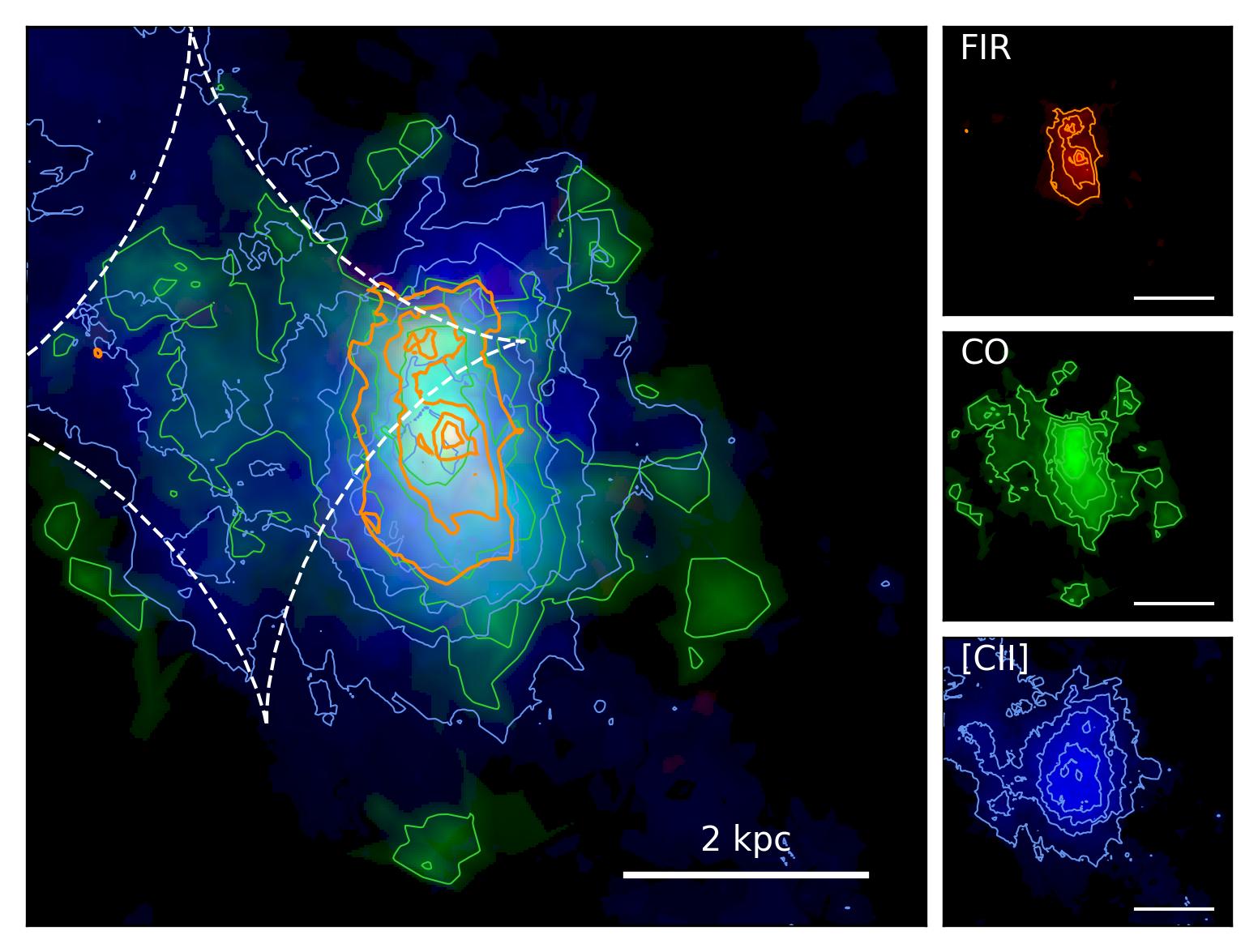}

\caption{Composite image of the FIR, CO and [\ion{C}{ii}] emission in SDP.81. The contours are drawn at 20, 40, 60 and 80~per cent of the FIR/CO/[\ion{C}{ii}] surface brightness maximum. The CO map is made by adding the CO(3--2), (5--4), (8--7) and (10--9) maps in units of $L_\odot$. \label{fig:composite}}
\end{centering}
\end{figure*}

\subsubsection{ALMA Band 8 continuum}

The Band 8 continuum emission is observed at the highest S/N; consequently, the small-scale structure can be robustly reconstructed. Similarly to the Bands~6 and 7 continuum, the Band~8 continuum surface-brightness distribution is concentrated into two prominent clumps, corresponding roughly to the northern and central parts of the source, surrounded by a fainter emission approximately $3\times2$~kpc in extent (R15a,\citealt{Dye2015}). The northern clump is quadruply imaged, which results in a superb S/N and reconstruction fidelity, while the central clump is only doubly imaged. The relative brightness of the clumps is roughly constant across Bands 6, 7 and 8, indicating only a limited change in the dust spectral energy distribution (Section \ref{subsec:dust_sed}). 

\subsubsection{[CII] emission}

Thanks to the high S/N of the [\ion{C}{ii}] observations, we are able to model the full velocity structure of the [\ion{C}{ii}] emission, rather than just the frequency-averaged line. We frequency-averaged the line into channels 62.50~MHz ($\sim40$~km s$^{-1}$) wide, which provide a good trade-off between S/N per channel (necessary for a robust source reconstruction) and resolving the velocity structure; we then optimize for the source regularization parameter $\lambda_S$ for each channel separately. The velocity resolution matches that of the CO(5--4) and CO(8--7) reconstructions from R15b. We calculate the velocity moment-zero (surface brightness) and moment-one maps. For the moment-one map, we mask pixels with SNR$\leq$1 in each channel before calculating the velocity map.

The [\ion{C}{ii}] emission is considerably more extended than both the FIR continuum and the CO(5--4) and CO(8--7) emission (Fig.~\ref{fig:reconstructions}). Although the [\ion{C}{ii}] luminosity increases in the FIR-bright region of the source, at the position of the peak FIR surface brightness, the [\ion{C}{ii}] emission shows a significant drop, indicating an extremely low [\ion{C}{ii}]/FIR ratio. A similar drop in [\ion{C}{ii}]/FIR ratio at the continuum peak was recently found in a $z=1.7$ lensed DSFG SDP.11 \citep{Lamarche2018}, in which the regions with the highest $\Sigma_\mathrm{SFR}$ are essentially undetected in the [\ion{C}{ii}] line emission. We will address the [\ion{C}{ii}]/FIR ratio in SDP.81 in more detail in Section~\ref{subsec:CII_FIR}.

\begin{figure*}
\begin{centering}
\includegraphics[height=8 cm, clip=true]{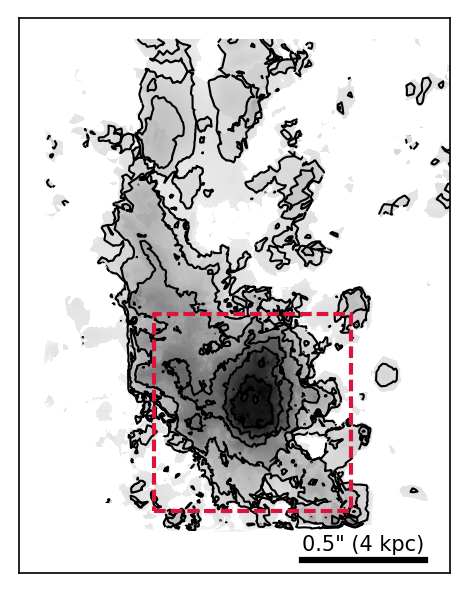}
\includegraphics[height=8 cm, clip=true]{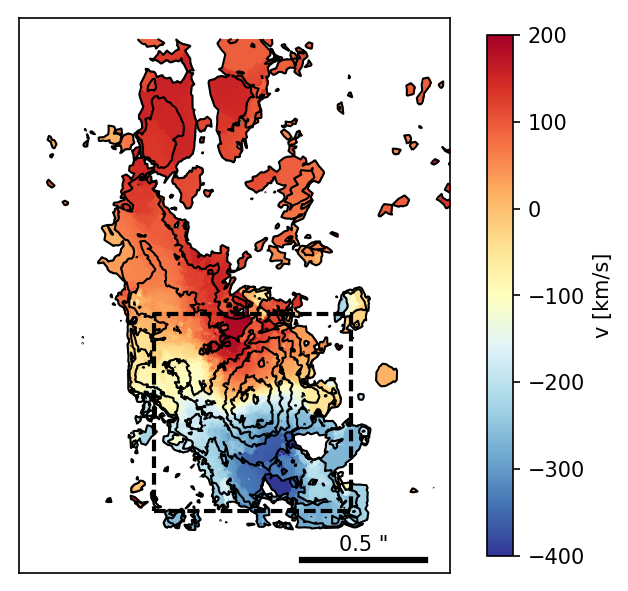}

\caption{[\ion{C}{ii}] surface brightness (moment-zero) and velocity (moment-one) maps, obtained from channels maps from Appendix~\ref{sec:appendix_A}. The [\ion{C}{ii}] emission extends some 10~kpc to the north of SDP.81. The black contours trace the moment-zero (surface brightness) at 5, 10, 20, 40, 60 and 80~per cent of the peak [\ion{C}{ii}] surface brightness. For the moment-one map calculation, we discard pixels with SNR$<$2 in individual channels. The dashed rectangles indicate the field of view shown in Fig.~\ref{fig:reconstructions}. We use the radio definition of the velocity in the LSRK frame and a systemic redshift $z=3.042$. \label{fig:cii_mom1}}
\end{centering}
\end{figure*}

The reconstructed [\ion{C}{ii}] emission map reveals a low surface-brightness emission ($\Sigma_\mathrm{[CII]}\sim3\times10^7~L_\odot$ kpc$^{-2}$) between -20 and +170~km s$^{-1}$, stretching out to $\sim$10~kpc north of SDP.81 (Fig.~\ref{fig:cii_mom1}). The velocity maps show that this excess emission is not co-rotating with the main [\ion{C}{ii}]/CO(5--4)/CO(8--7) component. This double-imaged feature is present in the dirty images of the corresponding velocity channels (Fig.~\ref{fig:reconstructions_appendix_cii}), clearly offset from the main Einstein arc. The total $L_\mathrm{[CII]}$ luminosity of this component is $\sim0.6\times10^9~L_\odot$, less than 10~per cent of the total source-plane [\ion{C}{ii}] luminosity. We do not detect any FIR continuum or CO emission from the region outside the dashed box in Fig.~\ref{fig:cii_mom1}, despite the high S/N of the FIR and CO(5--4)/CO(8--7) data. This second [\ion{C}{ii}] source is not co-spatial with the optically-bright substructure detected in the HST WFC3 160W imaging (\citealt{Dye2014}, R15b). Given the perturbed velocity/velocity-dispersion fields (R15b), we interpret this morphology as a potential low-mass companion; assuming the $M_\mathrm{gas}$ scales with $L_\mathrm{[CII]}$, the gas mass ratio of the two components is 10:1. Assuming this [\ion{C}{ii}] emission is due to star-formation, we use the \citet{Herrera2015} SFR-[\ion{C}{ii}] relation to derive a typical $\Sigma_\mathrm{SFR}=1$~M$_\odot$ yr$^{-1}$. It total, $\sim$50~per cent of the total [\ion{C}{ii}] luminosity arises from regions with $\Sigma_\mathrm{SFR}\leq10~M_\odot$~yr$^{-1}$.

\subsubsection{CO(3--2) emission}
The CO(3--2) emission is only detected in the northern part of the FIR-bright region in SDP.81, at $\sim5\sigma$ significance. As the CO(3--2) has much lower S/N ($\sim10$) than the remaining lines/continuum data, it must be interpreted carefully. A key check on the relative source-plane distribution of individual tracers is to compare their image-plane surface brightness distributions. As seen in Fig.~\ref{fig:cleaned}, in the image plane, the CO(3--2) emission peaks in a different region compared to the Band~8 continuum and [\ion{C}{ii}]. We confirm that this is not an artifact of a sparse $uv$-plane coverage by reconstructing mock ALMA observations matching the S/N and $uv$-plane coverage of the CO(3--2) data (Appendix~\ref{sec:appendix_B}): for a uniformly distributed CO(3--2) emission, emission from the southern part of the counterimage would be detected. This confirms that the CO(3--2) emission is concentrated to the north of the source. In the sky-plane, the CO(1--0) emission peaks to the south of the FIR source (\citealt{Valtchanov2011}, R15b); which implies a significant offset between the CO(3--2) and CO(1--0) surface brightness peaks. 

\subsubsection{CO(10--9) emission}
The CO(10--9) emission is concentrated between the diamond caustics, and to the north, continuing the trend seen in the CO(8--7) emission. The lower S/N makes the interpretation of the faint extended structure (surface brightness <1.2~mJy km~s$^{-1}$ kpc$^{-2}$) challenging: as we show in Appendix~\ref{sec:appendix_C}, this structure can be an artifact of the lensing reconstruction.

Finally, Fig.~\ref{fig:composite} shows the relative distribution of the FIR continuum, [\ion{C}{ii}] and CO emission. This highlights the compact nature of the FIR continuum with respect to the CO and [\ion{C}{ii}] lines, and the extended, low-surface brightness [\ion{C}{ii}] emission.

\section{Results and discussion}
\label{sec:results+discussion}

\subsection{Dust continuum}
\label{subsec:dust_sed}

\subsubsection{Global dust SED}

First, we consider the dust thermal spectral energy distribution (SED). Namely we use the ALMA Bands 4/6/7/8 and \textit{Herschel} PACS/SPIRE observations listed in Table~\ref{tab:sed} and an optically-thin modified black-body SED:

\begin{equation}
S_\nu \propto M_\mathrm{dust} \frac{\nu^{3+\beta}}{\exp(-h\nu/k T_\mathrm{dust})-1},
\label{eq:mbb}
\end{equation}

where $M_\mathrm{dust}$ is the dust mass, $h$ is the Planck constant, $k$ is the Boltzman constant, $T_\mathrm{dust}$ the dust temperature and $\beta$ the emissivity index. As the continuum magnification varies only marginally between ALMA Bands 6/7/8, we assume that the (unresolved) \textit{Herschel} PACS/SPIRE continuum follows the same source-plane surface brightness distribution as ALMA Band~8 continuum, and hence has the same magnification factor. We assume a uniform $T_\mathrm{dust}$ and $\beta$, and adopt a uniform magnification for the 160-$\mu$m to 1.3-mm continuum. We obtain $T_\mathrm{dust}=35\pm4$~K and $\beta=1.8\pm0.3$ and a (magnification-corrected) FIR luminosity (8-1000~$\mu$m) $L_\mathrm{FIR}=(2.4\pm0.5)\times10^{12}$~L$_\odot$. These are in agreement with previous estimates (\citealt{Bussmann2013, Swinbank2015, Sharda2018}; note that \citealt{Sharda2018}, do not explicitly fit for $\beta$). The inferred $T_\mathrm{dust}$, $\beta$ are in line with those derived for the general high-redshift DSFG population. For example, for the 99 DSFGs from the ALESS sample \citep{Hodge2013} \citet{Swinbank2014} derived a median $T_\mathrm{dust}=35\pm1$~K using \textit{Herschel} SPIRE and ALMA Band~7 photometry (approach directly comparable to ours); while \citet{daCunha2015} found a median $T_\mathrm{dust}=43\pm10$~K by including additional UV- to radio-wavelength photometry. The dust emissivity index $\beta$ is within the range of $\beta=1.5-2.0$ inferred for ALESS galaxies \citep{Swinbank2014}.

Although the increasing CMB temperature at high redshifts can significantly bias the inferred $T_\mathrm{dust}$ and $L_\mathrm{FIR}$ \citep{daCunha2013, Zhang2016}, we consider the CMB effects to be negligible for SDP.81, Namely, adopting the \citet{daCunha2013} corrections, the CMB effect on $T_\mathrm{dust}$ is $\leq0.1$~K, while the inferred $L_\mathrm{FIR}$ is biased by $\leq$0.1~dex.

\begin{figure}
\begin{centering}
\includegraphics[width=7 cm, clip=true]{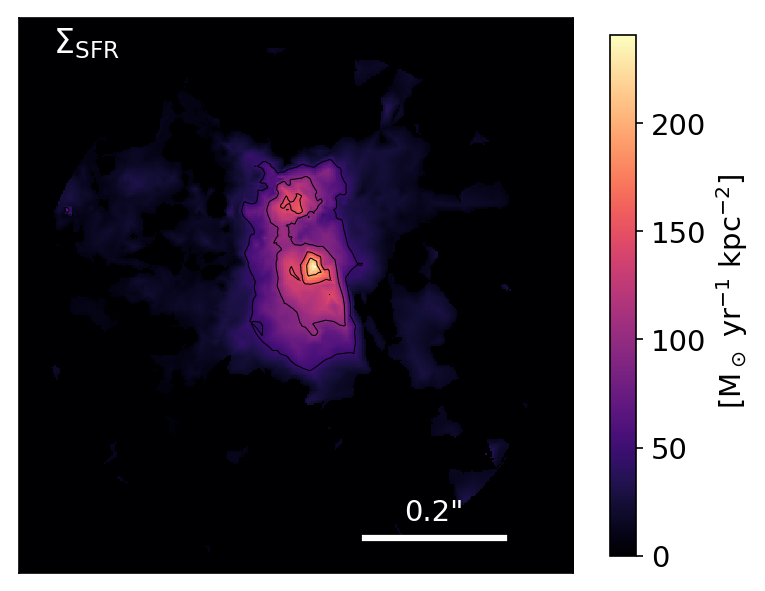}

\caption{Star-formation rate surface density map, obtained by combining the reconstructed ALMA Bands 6, 7 and 8 continuum maps, and assuming a Salpeter IMF. The contours start at 50, 100, 150 and 200 ~M$_\odot$ yr$^{-1}$ kpc$^{-2}$. \label{fig:sigma_SFR}}
\end{centering}
\end{figure}

We derive the global star-formation rate and the resolved star-formation rate surface density $\Sigma_\mathrm{SFR}$ (Fig.~\ref{fig:sigma_SFR}) using the \citet{Kennicutt1998} SFR-$L_\mathrm{FIR}$ relation for the Salpeter initial mass function (see also \citealt{Casey2014}):
\begin{equation}
\mathrm{SFR} [M_\odot~\mathrm{yr}^{-1}] = 1.71\times10^{-10} L_\mathrm{FIR} [L_\odot],
\end{equation}
which yields SFR = 410$\pm$90~M$_\odot$ yr$^{-1}$. Note that a part of the FIR luminosity in intensely star-forming systems might be due to FUV heating contribution from older stellar populations \citep{Narayanan2015}; however, this effect only becomes important at SFRs much higher than that of SDP.81. Given the compact size of the FIR emission in SDP.81 compared to the CO and [\ion{C}{ii}] emission, we will refer to the region with $\Sigma_\mathrm{SFR}\geq 50$~M$_\odot$ yr$^{-1}$ kpc$^{-2}$ as the \emph{FIR-bright region}.

\begin{table}
 \caption{FIR and mm-wave photometry of SDP.81 used for the dust SED modelling, with \textit{Herschel} PACS and SPIRE flux-densities adopted from \citet{Zhang2018b} and ALMA Bands 6 and 7 flux-densities adopted from R15a. The \textit{Herschel} flux measurements are given in the image-plane; for our analysis, we assume all the 160 to 500-$\mu$m emission to follow the Band~8 surface brightness distribution.}
 \label{tab:sed}
 \centering
  \begin{tabular}{@{} l|c @{}}
  \hline
  Band & [mJy] \\
  \hline
  PACS 160~$\mu$m & (58$\pm$10)/$\mu$\\
  SPIRE 250~$\mu$m & (133$\pm$11)/$\mu$\\
  SPIRE 350~$\mu$m & (186$\pm$14)/$\mu$\\
  SPIRE 500~$\mu$m & (165$\pm$14)/$\mu$\\
  ALMA Band 8 (640~$\mu$m) & 5.0$\pm$0.5 \\
  ALMA Band 7 (1.0~m) & 1.9$\pm$0.2 \\
  ALMA Band 6 (1.3~mm) & 1.14$\pm$0.11\\
 \hline
 \end{tabular}
 
\end{table}

\subsubsection{Resolved dust SED}

Having derived the spatially-integrated $T_\mathrm{dust}$ and $\beta$, we now consider the spatial variation in the dust continuum SED using the source-plane reconstructions of the ALMA Bands 6, 7 and 8 continuum (rest-frame wavelength 160--350~$\mu$m). These are well within the Rayleigh-Jeans tail of the modified black-body profile, and hence can not robustly constrain both $T_\mathrm{dust}$ and $\beta$ at the same time. We therefore adopt $\beta=1.8$ derived from the spatially-integrated SED fit, and fit for $T_\mathrm{dust}$ only.

However, fitting the modified black-body SED to the resolved Bands 6, 7 and 8 data, we do not find any significant spatial variation in $T_\mathrm{dust}$; the variation in $\Sigma_\mathrm{FIR}$ is fully accounted by a varying $M_\mathrm{dust}$. Repeating the SED-fitting on the {\sc Clean}ed Bands 6, 7 and 8 images (thus eliminating any lens-modelling systematics) yields results consistent with the source-plane analysis. Therefore, for the rest of this paper, we adopt a uniform $T_\mathrm{dust}$ approximation for the FIR luminosity calculation.

The uniform dust temperature in SDP.81 contrasts with the expectation of $T_\mathrm{dust}$ increasing towards the centre of DSFGs; for example, \citet{CR18} invoked a radial decrease in $T_\mathrm{dust}$ and dust optical depth to explain the different FIR and CO(3--2) sizes in the stacked observations of the ALESS~sample, with $T_\mathrm{dust} \geq 80$~K in the central $R \leq 1$~kpc region (A.~Wei\ss, private communication). 

\begin{figure}
\begin{centering}
\includegraphics[height=5cm, clip=true]{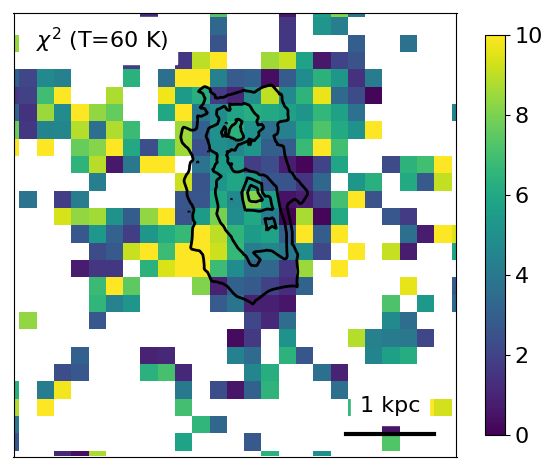}
\includegraphics[height=5cm, clip=true]{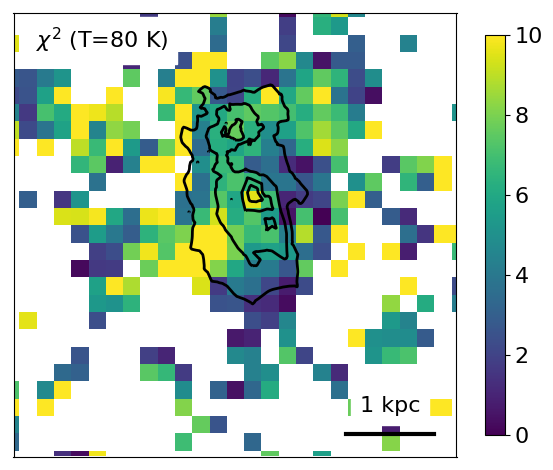}

\caption{Constraints on dust temperature in SDP.81: reduced $\chi^2$ for $T_\mathrm{dust}=60$~K (right) and 80~K (left). In contrast to temperature gradient models of \citet{CR18}, the FIR-bright clumps are inconsistent with $T_\mathrm{dust}\geq 80$~K (reduced $\chi^2\geq5$). \label{fig:t_dust}}
\end{centering}
\end{figure}

We therefore consider whether our data might support high $T_\mathrm{dust}$ ($\geq$60~K) in the central FIR-bright clumps. We do this by fixing $T_\mathrm{dust}$ to 60 and 80~K, respectively, and optimizing for the normalization ($\propto M_\mathrm{dust}$) only. We find $T_\mathrm{dust}$=60~K to be inconsistent with the data at $\geq3\sigma$ significance anywhere except in the vicinity of the FIR brightness maximum (Fig.~\ref{fig:t_dust}); $T_\mathrm{dust}$=80~K is excluded at $\geq3\sigma$ confidence level for the entire source. These results are therefore in tension with the high central $T_\mathrm{dust}$ invoked by \citet{CR18}.

Finally, we note three fundamental limitations of inferring $T_\mathrm{dust}$ from the rest-frame 160--350~$\mu$m continuum. First, at 200-pc scales probed here, the high-$T_\mathrm{dust}$ regions might not dominate the continuum SED, resulted in a bias towards low $T_\mathrm{dust}$ temperature estimates due to the mass-weighting in Equation~(\ref{eq:mbb}). Second, our resolved ALMA imaging is still on the Rayleigh-Jeans tail of the dust continuum SED; higher-frequency, resolved observations (e.g., with ALMA Bands~9 or 10) will be necessary to properly constrain $T_\mathrm{dust}$ and $\beta$ on sub-kpc scales. Finally, the optically thin approximation make break down in the central regions, biasing the inferred $T_\mathrm{dust}$.

\subsection{Emission lines}
\label{subsec:gas_cooling}

With the resolved maps of the [\ion{C}{ii}] line and the almost complete CO SLED in hand, we can compare the total gas cooling via the [\ion{C}{ii}], [\ion{O}{i}] and all the CO lines combined. In particular, for the strong FUV fields and high gas (surface) densities expected in DSFGs, as the outer C$^+$ layer becomes progressively thinner, the main cooling channel of the neutral gas can switch from the [\ion{C}{ii}] line to the [\ion{O}{i}] 63~$\mu$m/145$\mu$m fine-structure lines, or even CO rotational lines \citep{Kaufman1999, Kaufman2006, Narayanan2017}.

Table~\ref{tab:co_cii_balance} lists the inferred source-plane CO and [\ion{C}{ii}] line luminosities, integrated over the entire source. We complemented the resolved ALMA observations by archival observations of the CO(1--0) (VLA; \citealt{Valtchanov2011}) and CO(7--6) (CARMA, \citealt{Valtchanov2011}) lines, as well as an upper limit\footnote{Unlike \citet{Lupu2012} who report a marginal (2$\sigma$) detection of the CO(9--8) line, we consider the CO(9--8) line to be undetected. We adopt a $3\sigma$ upper limit of $0.49\times10^9$~L$_\odot$ for the CO(9-8) line luminosity.} on the CO(9--8) lines (Z-Spec on the Caltech Submillimeter Observatory, \citealt{Lupu2012}). The unresolved/low-resolution nature of the archival data prevents a source-plane reconstruction. Therefore, we assume that the CO(1--0), (7--6) and (9--8) lines follow the same spatial distribution as the CO(3--2), (8--7) and (10--9) lines, respectively.

\begin{table*}
\caption{Spatially integrated [\ion{C}{ii}] and CO line luminosities, with corresponding lensing magnifications $\mu$. We list the source-plane values inferred in this work and Rybak et al., (2015b; R15b), along with CO(1--0), (7--6) and (9--8) luminosities from \citet{Valtchanov2011} and \citet{Lupu2012}. To infer the source-plane line luminosities, we assume that CO(1--0) has the same spatial distribution as CO(3--2), CO(7--6) as CO(8--7), and CO(9--8) as CO(10--9). \label{tab:co_cii_balance}}
\begin{center}
 \begin{tabular}{@{}l|cccc @{}}
 \hline \hline
Line & $\mu$ & $L_\mathrm{line}$ & $L'_\mathrm{line}$ & Ref. \\
 & & [$10^{9}$ L$_\odot$] & [$10^{10}$ K km s$^{-1}$ pc$^{-2}$] & \\
\hline
FIR & 18.2$\pm$1.2 & 2400$\pm$500 & -- & R15a (Bands 6 \& 7) \\
 & & & & This work (Band~8) \\
\ion{C}{ii} & 15.3$\pm$0.5 & 4.36$\pm$0.44 & 1.99$\pm$0.20 & This work\\
\ion{O}{i} 63~$\mu$m & -- & 2.5$\pm$0.7 & 1.4$\pm$0.4 & \citet{Zhang2018b} \\ 
CO(1--0)$^a$ & --- & 0.005$\pm$0.001 & 11$\pm$2 & Valtchanov et al., (2011)\\
CO(3--2) & 17.0$\pm$0.4 & 0.018$\pm$0.010 & 1.33$\pm$0.13 & This work\\
CO(5--4) & 17.9$\pm$0.7 & 0.075$\pm$0.024 & 1.23$\pm$0.12 & R15b\\
CO(7--6)$^a$ & -- & 0.06$\pm$0.02 & 0.6$\pm$0.2 & Valtchanov et al., (2011)\\
CO(8--7) & 16.9$\pm$1.1 & 0.098$\pm$0.020 & 0.39$\pm$0.04 & R15b \\
CO(9--8)$^a$ & -- & $<$0.49 & $<$0.21 & Lupu et al. (2012)\\
CO(10--9) & 20.6$\pm$2.6 & 0.0060$\pm$0.0010 & 0.0124$\pm$0.0012 & This work\\
\hline
 \end{tabular}
 \end{center}
 
 $^a$ inferred from unresolved observations.
\end{table*}

\subsubsection{[CII] dominates the gas cooling budget}

We first compare the source-averaged [\ion{C}{ii}], [\ion{O}{i}] and CO SLED cooling rates.
The [\ion{O}{i}] 63-$\mu$m line is detected in \textit{Herschel} SPIRE FTS spectra \citep{Valtchanov2011, Zhang2018b} with a line flux of $(2.1\pm0.7)\times10^{-18}$~W~m$^{-2}$, while the 3$\sigma$ upper limit on the [\ion{O}{i}]~145-$\mu$m line flux is $2.3\times10^{-18}$~W~m$^{-2}$ \citep{Zhang2018b}. This translates to image-plane line luminosities of $\mu L_\mathrm{[OI]63}=(45\pm12)\times10^{9}$~L$_\odot$, $\mu L_\mathrm{[OI]145}<5\times10^{10}$~L$_\odot$. Assuming both lines follow the FIR continuum surface brightness distribution, we obtain the de-lensed source-plane luminosities of $L_\mathrm{[OI]63}=(2.5\pm0.7)\times10^9$~L$_\odot$ and $L_\mathrm{[OI]145}<3\times10^9$~L$_\odot$, respectively. The inferred [\ion{O}{i}]/[\ion{C}{ii}] luminosity ratio is 0.6$\pm$0.2 for the [\ion{O}{i}] 63-$\mu$m and $\leq$0.7 (but likely much lower) for the [\ion{O}{i}] 145-$\mu$m line.

To compare the total gas cooling via the CO ladder and the [\ion{C}{ii}], we first obtain a first-order estimate of the total cooling via the CO $J_\mathrm{upp}=1-10$ lines (in the absence of CO(2--1), CO(4--3), CO(6--5) and CO(7--6) observations) assuming that the CO line ratios in SDP.81 match CO ratios of the Class~I ULIRGs from \citet{Rosenberg2015}. Given the low CO(10--9) luminosity, the $J_\mathrm{upp}>10$ CO rotational lines are expected to contribute only marginally to the gas cooling. The total CO luminosity is $L_\mathrm{CO}=\Sigma_{J=1}^{10} L_\mathrm{CO(J_\mathrm{upp})}\simeq0.8\times10^9~L_\odot$, $\sim$20~per cent of the total $L_\mathrm{[CII]}$ and $\sim$10~per cent of the total gas cooling ($L_\mathrm{[CII]}+L_\mathrm{CO}+L_\mathrm{[OI]}$). Therefore, the global gas cooling in SDP.81 is dominated by the [\ion{C}{ii}] line.

\begin{figure}
\begin{centering}
\includegraphics[height=5.5cm, clip=true]{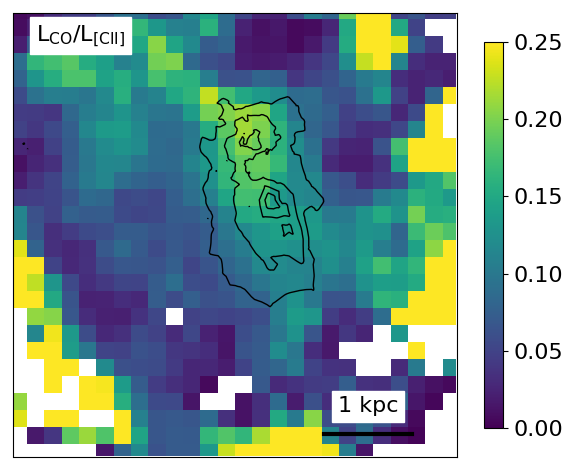}\\

\caption{Resolved $L_\mathrm{CO}$/$L_\mathrm{[CII]}$ ratio, assuming a \citet{Rosenberg2015} Class~I ULIRG CO SLED to estimate the full CO excitation. The [\ion{C}{ii}] line dominates the gas cooling across the entire source; the highest CO/[\ion{C}{ii}] ratio ($\sim0.25$) is found around the northern FIR-bright clump. \label{fig:gas_cooling}}
\end{centering}
\end{figure}

Fig.~\ref{fig:gas_cooling} shows the resolved CO/[\ion{C}{ii}] gas cooling ratio. The [\ion{C}{ii}] line dominates over the CO cooling across the entire source, with the highest CO/[\ion{C}{ii}] cooling ratio ($\sim20$~per cent) found at the position of the northern star-forming clump. The low CO/[\ion{C}{ii}] cooling ratios indicate molecular cloud surface density $\leq10^3$~M$_\odot$ pc$^{-2}$ (c.f.,\,\citealt{Narayanan2017}). 

\subsubsection{CO spectral energy distribution}

\begin{figure}
\begin{centering}
\includegraphics[width=8cm, clip=true]{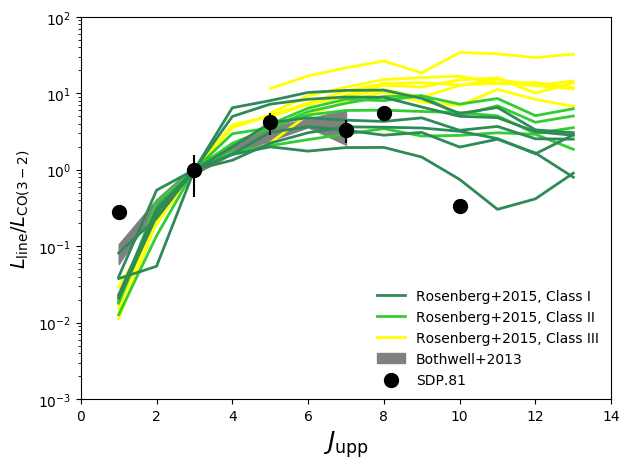}\\
\includegraphics[width=8cm, clip=true]{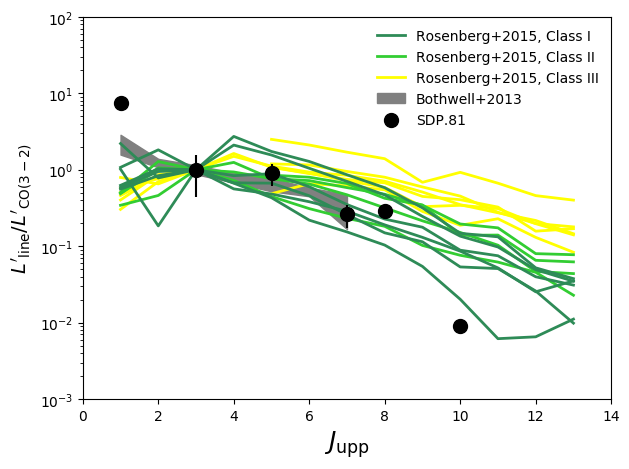}\\

\caption{Source-plane CO SLED in SDP.81 normalized to the CO(3--2) luminosity in units of L$_\odot$ (\textit{upper}) and in K km s$^{-1}$ pc$^2$ (\textit{lower}), compared to the ULIRGs from the \citet{Rosenberg2015} sample, colour-coded by the ULIRG "class'' (I -- dark green, II -- green, III -- yellow; only ULIRGs with CO(3--2) detections are considered.), and the \citep{Bothwell2013} CO SLED survey in $z\geq1$ DSFGs. The CO SLED excitation in SDP.81 matches with that of the \citet{Rosenberg2015} Class~I ULIRGS, which are fully consistent with UV-only heating. \label{fig:co_sled}}

\end{centering}
\end{figure}

The global source-plane CO SLED in SDP.81 peaks between $J_\mathrm{upp}=5$ and 9 (in units of $L_\odot$), with a sharp drop at $J_\mathrm{upp}=10$ (Fig.~\ref{fig:co_sled}, see also \citealt{Yang2017}, for a recent sky-plane analysis). Compared to the \citet{Rosenberg2015} study of CO excitation in present-day ULIRGS, the sharp drop in $L_\mathrm{CO}$ for $J_\mathrm{upp}\geq9$ and minor CO contribution to the global neutral gas cooling makes SDP.81 akin to \citet{Rosenberg2015} Class~I sources, which have similarly low CO cooling contributions ($\leq10$~per cent).
In addition to the heating by FUV radiation, mechanical heating by dissipating supernovae shocks can contribute significantly to the gas energy budget in starburst regions (e.g., \citealt{Meijerink2011, Kazandjian2012, Kazandjian2015}), pumping the $J_\mathrm{upp}\geq10$ CO transitions, as well as high-$J$ HCN, HNC and $^{13}$CO lines. Significant mechanical heating is required to explain the high-$J$ CO excitation in some of the archetypal present-day ULIRGs, such as e.g., Arp~220 \citep{Rangwala2011}, NGC~253 \citep{Rosenberg2014} and NGC~6420 \citep{Meijerink2013}. Given the absence of CO $J_\mathrm{upp}>10$, HCN or $^{13}$CO observations in SDP.81, we do not attempt to directly constrain the mechanical heating contribution. 
As the SDP.81 CO ladder is well-reproduced ($\chi^2\leq1$) by the {\sc PDRToolbox} models which do not include mechanical heating and as the SDP.81-like \citet{Rosenberg2015} Class~I ULIRGs are fully compatible with UV heating, the observed line ratios in SDP.81 are consistent with negligible mechanical heating.

Finally, we consider how closely individual CO lines trace the FIR emission. We calculate the Pearson's correlation coefficient $R$ for each line and estimate the corresponding uncertainty by drawing 1000 random realizations of CO/[\ion{C}{ii}] and FIR surface brightness (with the mean and standard deviation from Fig.~\ref{fig:reconstructions}) for each pixel with SNR$\geq$3 in both tracers. We find $R_\mathrm{CO(3-2)/FIR}=0.401\pm0.010$, $R_\mathrm{CO(5-4)/FIR}=0.847\pm0.006$, $R_\mathrm{CO(8-7)/FIR}=0.821\pm0.006$ and $R_\mathrm{CO(10-9)/FIR}=0.664\pm0.09$. However, the stronger regularization of CO reconstructions compared to the FIR continuum might artificially decrease the CO--FIR correlation by smoothing the small-scale CO features, especially at high $\Sigma_\mathrm{FIR}$. For comparison, the correlation coefficient for [\ion{C}{ii}] $R_\mathrm{[CII]/FIR}=0.712\pm0.006$

\subsubsection{Gas depletion time across the source}
\label{subsec:t_dep}

Finally, we combine the resolved $\Sigma_\mathrm{SFR}$ maps and CO emission line maps to estimate the molecular gas depletion time, $t_\mathrm{dep}\sim\Sigma_\mathrm{H_2}/\Sigma_\mathrm{SFR}$. We estimate the molecular gas surface density $\Sigma_\mathrm{H_2}$ using the CO(3--2) line map. For $J_\mathrm{upp}>1$ CO transitions, the $L'_\mathrm{CO}$ is first down-converted to $L'_\mathrm{CO(1-0)}$ via the $R_{J1} = L'_\mathrm{CO(J-J-1)}/L'_\mathrm{CO(1-0)}$ factor, which is then converted to $M_\mathrm{gas}$ via $M_\mathrm{gas}=\alpha_\mathrm{CO} L'_\mathrm{CO(1-0)}$. We adopt $R_{31}$ from \citet{Sharon2016} who found a mean $R_{31}=0.78\pm0.27$ for a sample of $z\sim2$ sub-millimeter galaxies; and $\alpha_\mathrm{CO}=1.0$ inferred by \citet{CR18} from CO kinematic studies of $z=2-3$ DSFGs (a similar value has been derived by \citealt{Bothwell2013}). These are consistent with the dynamical mass constraints from the CO(5--4) and (8--7) velocity maps (R15b, \citealt{Swinbank2015}).

\begin{figure}
\begin{centering}
\includegraphics[height=5.5cm, clip=true]{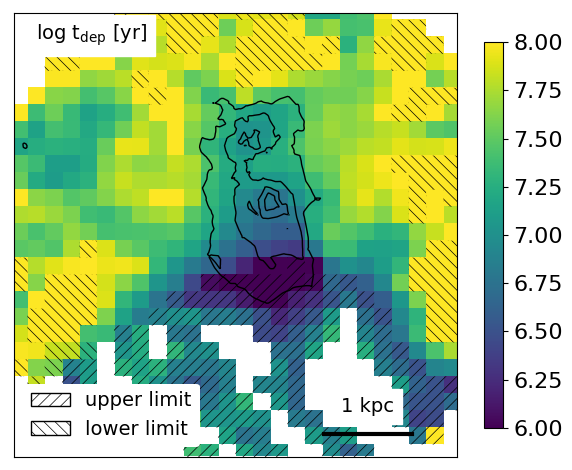}\\

\caption{Molecular gas depletion time $t_\mathrm{dep}$ across SDP.81; $t_\mathrm{dep}$ varies from $\sim100$~Myr at the outskirts of the source (with very little FIR emission), to $\sim1$~Myr in the southern part of the FIR-bright region. We denote the 3$\sigma$ upper limits on $t_\mathrm{dep}$ (regions with S/N$\geq3$ in FIR and $<3$ in CO(3--2)) and 3$\sigma$ lower limits (S/N $<3$ in FIR and $\geq3$ in CO(3--2)).  \label{fig:t_depletion}}
\end{centering}
\end{figure}

Fig.~\ref{fig:t_depletion} shows the resolved $t_\mathrm{dep}$ map in SDP.81. With the CO(3--2) emission concentrated to the north, we find a strong variation of $t_\mathrm{dep}$ across the FIR bright source, from $\sim1$~Myr in the south to $\sim20$~Myr in the north. By adopting the \citet{Sharon2016} $r_{3/1}$ value, the relative uncertainty is $\sim$33~per cent; the flux calibration uncertainty on the CO(3--2)/FIR ratio is $\sim15$~per cent.

Due to the low S/N and structure in the reconstructed CO(3--2) emission, we refrain from investigating the Kennicutt-Schmidt relation, particularly the slope of $\Sigma_\mathrm{H_2}-\Sigma_\mathrm{SFR}$ relation. However, as the concentration of the CO(3--2) emission to the north (and the low CO(3--2) surface brightness to the south) is robustly recovered by the lens modelling (Appendix~\ref{sec:appendix_B}), the general $t_\mathrm{dep}$ trend is robust. As a further check, using the CO(5--4) line as a molecular gas tracer\footnote{although CO(5--4) generally correlates closely with FIR continuum \citep{Daddi2015}, and hence does not directly trace $t_\mathrm{dep}$.} and adopting $R_{51}=0.32\pm0.05$ from the \citet{Bothwell2013} CO SLED survey of a sample of 40 DSFGs at $z=1.2-4.1$, we find a $t_\mathrm{dep}=1-30$~Myr, confirming the extremely short gas depletion time.

Resolved $t_\mathrm{dep}$ measurements have been carried out for only a handful of DSFGs (see \citealt{Sharon2019} for a recent compilation), which generally show similarly strong ($\sim1$~dex) variation in $t_\mathrm{dep}$. The depletion times in SDP.81 are in the extreme starburst regime, an order of magnitude lower than seen in some other high-redshift DSFGs -- e.g., GN20 ($t_\mathrm{dep}=50-200$~Myr, \citealt{Hodge2015}), J14011 (50-100~Myr, \citealt{Sharon2013}) and J0911 (200-1000~Myr, \citealt{Sharon2019}). \citet{Canameras2017} found a comparably low $t_\mathrm{dep}$ (10-100~Myr) in an extreme strongly lensed $z\sim3$ starburst PLCK~G244.8+54.9, albeit at much higher star-formation surface densities ($\Sigma_\mathrm{SFR}\sim2\times10^3$~M$_\odot$ yr$^{-1}$ kpc$^{-2}$). Given the extremely short depletion time in the centre of SDP.81, sustaining the high $\Sigma_\mathrm{SFR}$ requires an efficient loss of gas angular momentum to feed the star formation.

\subsection{[CII]/FIR ratio and deficit}
\label{subsec:CII_FIR}

Studies of nearby star-forming galaxies (e.g., \citealt{Malhotra2001, Luhman2003}) have revealed that the [\ion{C}{ii}]/FIR ratio decreases with increasing FIR luminosity (i.e., SFR) -- the so-called [\ion{C}{ii}]/FIR deficit. This trend was confirmed by \textit{Herschel} FIR surveys of local main-sequence galaxies (e.g., \citealt{Smith2017}) and ULIRGs (e.g., \citealt{Diaz2013,Diaz2017}), as well as by unresolved observations of high-redshift DSFGs (e.g., \citealt{Stacey2010,Gullberg2015, spilker2016}). In particular, resolved observations at $z\sim0$ have found a tight correlation with the FIR (SFR) surface density \citep{Herrera2015}. Recently, resolved [\ion{C}{ii}] studies in $z>1$ DSFGs have been presented by \citet{Gullberg2018, Lamarche2018, Litke2019} and \citet{Rybak2019}, revealing a pronounced [\ion{C}{ii}]/FIR deficit down to $L_\mathrm{[CII]}/L_\mathrm{FIR}\simeq10^{-4}$. Similarly pronounced [\ion{C}{ii}]/FIR deficits have been found in resolved ALMA observations of $z\geq5$ quasar hosts (e.g., \citealt{Wang2019, Neeleman2019, Venemans2019}). Here, we consider both the source-averaged and resolved [\ion{C}{ii}]/FIR ratio in SDP.81, and compare it to local and high-redshift studies.

Based on the source-plane FIR and [\ion{C}{ii}] luminosities (Table~\ref{tab:co_cii_balance}), we infer a spatially integrated [\ion{C}{ii}]/FIR luminosity ratio of $(1.5 \pm 0.3)\times 10^{-3}$. This is consistent with low [\ion{C}{ii}]/FIR ratios seen in other high-redshift sources (Fig.~\ref{fig:cii_fir_comparison}), as well as the \textit{Herschel} observations of ULIRGs from the GOALS sample \citep{Diaz2017}. However, as the [\ion{C}{ii}] emission is significantly more extended than the FIR continuum (in line with \citealt{Gullberg2018,Rybak2019}), the source-averaged [\ion{C}{ii}/FIR measurement provides only an upper limit on the [\ion{C}{ii}]/FIR ratio in the central, FIR-bright region. The large extent of the [\ion{C}{ii}] emission compared to the FIR continuum implies that the spatially-averaged [\ion{C}{ii}]/FIR measurements in high-redshift sources -- as inferred from unresolved observations -- mask the potentially very strong [\ion{C}{ii}]/FIR deficit in the actual star-forming regions. 

\begin{figure}
\begin{centering}

\includegraphics[height = 5.5cm, clip=true]{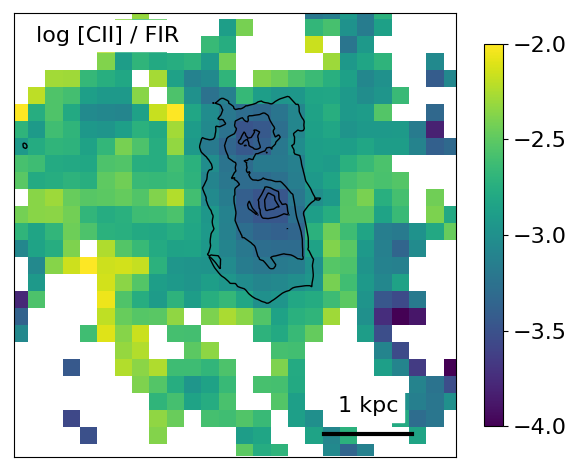}\\
\includegraphics[height = 5.5cm, clip=true]{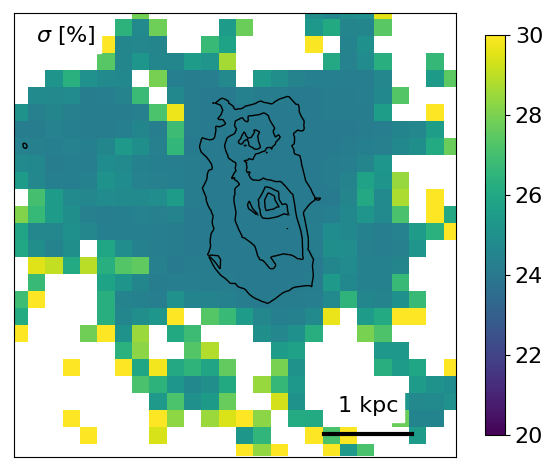}

\caption{200~pc mapping of the [\ion{C}{ii}]/FIR ratio in SDP.81 \label{fig:cii_fir_map}, with the associated fractional uncertainty. The [\ion{C}{ii}]/FIR varies by $>1.5$~dex on sub-kpc scales: while the outskirts of the system show efficient [\ion{C}{ii}] cooling ($L_\mathrm{[CII]}/L_\mathrm{[CII]}=10^{-2.5}-10^{-2.0}$, the FIR-bright star-forming clumps show a pronounced [\ion{C}{ii}]/FIR deficit, down to $10^{-3.5}$.} 
\end{centering}
\end{figure}

\begin{figure*}
\begin{centering}

\includegraphics[width=14 cm, clip=true]{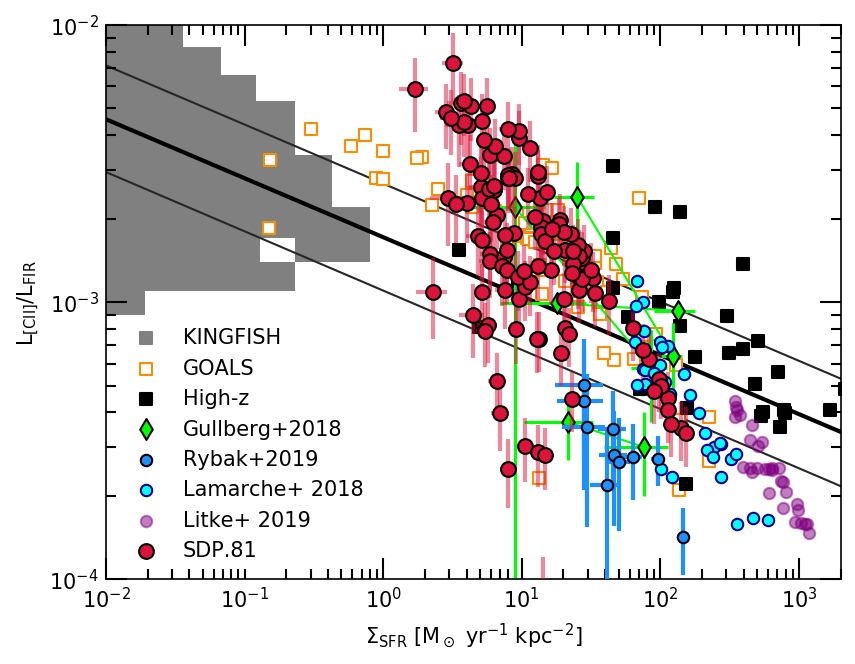}

\caption{Comparison of the resolved $L_\mathrm{[CII]}/L_\mathrm{IR}$ ratio in SDP.81 (red) with the other local (KINGFISH - \citealt{Smith2017}; GOALS - \citealt{Diaz2017}) and unresolved high-redshift observations (adopted from \citealt{Smith2017}); and resolved $z=2-6$ DSFG observations \citep{Gullberg2018, Lamarche2018, Litke2019, Rybak2019}. Only 1/4 of the datapoints from \citet{Litke2019} and SDP.81 are displayed for clarity. The bold lines indicate the empirical relation of \citet{Smith2017} with the associated 1$\sigma$ scatter. The outliers around [\ion{C}{ii}/FIR$\simeq2\times10^{-4}$, $\Sigma_\mathrm{SFR}$ are located at the very outskirts of SDP.81 (see Fig.~\ref{fig:cii_fir_map}) and are likely modelling artifacts. \label{fig:cii_fir_comparison}}
\end{centering}
\end{figure*}

Considering the resolved [\ion{C}{ii}]/FIR maps in Fig.~\ref{fig:cii_fir_map}, the [\ion{C}{ii}]/FIR ratio varies by more than 1~dex across the source, down to $\sim(2-3)\times10^{-4}$ in the two FIR-bright clumps -- almost an order of magnitude lower than the spatially-integrated [\ion{C}{ii}]/FIR ratio. We find significant ($\geq3\sigma$) gradients in the [\ion{C}{ii}]/FIR ratio on sub-kpc scales. Namely, on the outskirts, the [\ion{C}{ii}]/FIR ratio is consistent with expected gas heating efficiency via the photoelectric effect (0.1--1.0~per cent, e.g., \citealt{BakesTielens1994}). However, the [\ion{C}{ii}]/FIR ratio decreases steeply towards the FIR-bright clumps, down to $\sim3\times10^{-4}$, indicating a strong [\ion{C}{ii}]/FIR deficit. Similar radial trends have been observed in other resolved [\ion{C}{ii}]/FIR observations at high redshift \citep{Gullberg2015,Lamarche2018, Litke2019, Rybak2019, Wang2019}. As shown in Fig.~\ref{fig:gas_cooling}, the reduced efficiency of [\ion{C}{ii}] cooling is partially compensated by the increase in CO/[\ion{C}{ii}] luminosity ratio, as expected from radiative transfer arguments (e.g., \citealt{Narayanan2017}). We note that several isolated pixels at the very outskirts show very loiw [\ion{C}{ii}/FIR ratios - these are likely lens-modelling artifacts.

Fig.~\ref{fig:cii_fir_comparison} compares the [\ion{C}{ii}]/FIR ratio in SDP.81 as a function of the star-formation rate surface density $\Sigma_\mathrm{SFR}$ to other local and high-redshift sources, and the empirical relation of \citet{Smith2017}. Resolved [\ion{C}{ii}]/FIR deficit measurements in high-redshift DSFGs \citep{Gullberg2018,Lagache2018,Litke2019, Rybak2019} show a $\geq1$~dex scatter in [\ion{C}{ii}]/FIR for a given $\Sigma_\mathrm{SFR}$. While this might partially stem from systematic uncertainties in $\Sigma_\mathrm{SFR}$ estimates and the varying fraction of the [\ion{C}{ii}] emission from ionized gas (c.f.,\,\citealt{Diaz2017}), the observed [\ion{C}{ii}]/FIR dispersion is likely due to the intrinsic scatter in galactic properties, such as gas and dust conditions. From the SDP.81 data, we estimate the best-fitting $\log \Sigma_\mathrm{SFR}$-$\log$([\ion{C}{ii}]/FIR) slope of $-0.75\pm0.01$, using orthogonal distance regression, similar to the slope of $-0.7$ derived by \citet{Lamarche2018} for SDP.11. This is much steeper than the \citet{Smith2017} trend ($-0.21\pm0.03$). We address the [\ion{C}{ii}]/FIR slope when correcting for the fraction of gas in [\ion{C}{ii}] and the physical origin of the [\ion{C}{ii}]/FIR deficit in more detail in Section~\ref{subsec:cii_fir_origin}. 

\section{PDR modelling}
\label{sec:pdr_models}

\subsection{Modelling setup}
\label{sec:pdr_modelling_setup}
We adopt the line and FIR continuum predictions from the {\sc PDRToolbox} models \citep{Kaufman1999,Kaufman2006,Pound2008}. The {\sc PDRToolbox} models are calculated for a semi-infinite slab geometry, assuming a solar metallicity and stopping the calculation at the depth corresponding to a visual extinction $A_V=10$. The {\sc PDRToolbox} models have been widely used in interpreting atomic and molecular line observations from both $z\sim0$ ULIRGs (e.g., \citealt{Diaz2017}) and high-redshift DSFGs (e.g., \citealt{Valtchanov2011, Gullberg2015, Wardlow2017, Rybak2019}); we now apply the same approach to SDP.81. As the FIR studies of metallicity tracers in $z\geq1$ DSFGs indicate metallicities $\geq1~Z_\odot$ \citep{Wardlow2017}, we consider the $Z=1~Z_\odot$ default {\sc PDRToolbox} model as appropriate for dust-rich ISM of SDP.81. 

The default {\sc PDRToolbox} models are sampled onto a 0.25~dex grid which is too coarse to estimate the uncertainties on the inferred properties. We re-sample them onto a finer grid (steps of 0.05~dex), using a degree 2 spline for interpolation.

For the comparison with observations, we consider the following independent line ratios: [\ion{C}{ii}]/FIR, [\ion{C}{ii}]/CO(5--4), [\ion{C}{ii}]/CO(3--2), CO(8--7)/CO(5--4) and CO(10--9)/CO(5--4). The unresolved lines - [\ion{O}{ii}] and CO(7--6) are excluded from this comparison, as their magnification factors are unknown. Finally it should be noted that the [\ion{O}{i}] 63-$\mu$m line can undergo significant self-absorption (e.g., \citealt{Rosenberg2015}). The line ratios are calculated using line luminosities in units of $L_\odot$. In addition to the surface-brightness uncertainty due to lens modelling, we include an additional 10~per cent flux calibration error for each tracer, as appropriate for high-S/N ALMA observations. In the line-ratio maps, we consider pixels with SNR$\geq$3 as robust detections; for pixels with SNR$<3$, we adopt $3\sigma$ upper limits. We then calculate the $\chi^2$ for each $G$, $n$ model. Following \citet{Sawicki2012}, we include the $3\sigma$ upper limits as

\begin{multline}
\chi^2(G_i, n_i) = \sum_j \left( \frac{R^\mathrm{data}_j-R^\mathrm{model}_j(G_i, n_i) }{\sigma^\mathrm{data}_j} \right)^2 - \\
 - 2 \sum_{j'} \ln \left[ \sqrt{\frac{\pi}{2}} \sigma_{j'} \left( 1+\mathrm{erf}\left(\frac{R^\mathrm{model}_{j'}(G_i, n_i)-3\sigma_{j'} }{\sqrt{2} \sigma_{j'}} \right) \right) \right],
\label{eq:pdr_pf}
\end{multline}

where $R^\mathrm{data}_i$, $R^\mathrm{model}_i$ are the measured and predicted values of the $i$-th line ratio, the indices $j$, $j'$ run over the line ratios with SNR$\geq3$ (detections) and $<$3 (upper limits), respectively, and $\sigma_{j}$, $\sigma_{j'}$ denote corresponding uncertainties. Considering only the SNR$\geq$3 detections (i.e.\,dropping the second term in Equation~\ref{eq:pdr_pf}) changes the best-fitting $G$, $n$ only marginally ($\leq$0.1~dex).

Several corrections need to be applied to the data before a direct comparison with the PDR models. First, the observed [\ion{C}{ii}] luminosity has to be corrected for the contribution from non-PDR sources (i.e.\,ionized gas). The contribution of ionized gas to [\ion{C}{ii}] emission can be estimated using [\ion{N}{ii}] 122/205-$\mu$m lines \citep{Croxall2017, Herrera2018a}; however, [\ion{N}{ii}] 122-$\mu$m emission is undetected in the \textit{Herschel} spectra with only weak upper limits ($\leq2.6\times10^{-18}$~W m$^{-2}$ \citealt{Zhang2018b}, giving a PDR fraction of $\geq$0.6), while [\ion{N}{ii}] 205-$\mu$m line falls outside the \textit{Herschel} spectral coverage. Alternatively, using the empirical relation for the GOALS sample, the PDR contribution to the [\ion{C}{ii}] can be estimated using the rest-frame 63-$\mu$m and 158-$\mu$m continuum (Equation~4 of \citealt{Diaz2017}): for SDP.81, these correspond to the SPIRE 250-$\mu$m and ALMA Band~8 continuum (Table~\ref{tab:sed}), yielding $f_\mathrm{[CII]}^\mathrm{PDR}=0.90\pm0.10$. Given the available constraints, we adopt a conservative estimate of the $f_\mathrm{[CII]}^\mathrm{PDR}=0.8$. We discuss the sensitivity of the inferred PDR properties to this choice below.

Second, the {\sc PDRToolbox} predictions need to be adjusted in case a ratio of an optically thin and an optically thick tracer is considered. Namely, the emergent line intensities are predicted for a semi-infinite slab of gas illuminated only from the face side; for the extended starburst environment in SDP.81, the clouds are likely exposed to FUV radiation from multiple sides. While for the optically-thick tracers, only the emission from the face side of the cloud is observed, for the optically-thin tracers, both the sides facing to and away from the observer will contribute to the observed flux, and the predicted fluxes need to be doubled. The FIR continuum is typically optically thin down to rest-frame 70-100$\mu$m (e.g., \citealt{Casey2014}). We assume that the [\ion{C}{ii}] emission is optically thin, although moderate optical depths ($\tau\sim1$) have been proposed for some Galactic PDRs (e.g., \citealt{Graf2012, Sandell2015}) and high-redshift sources \citep{Gullberg2015}. The CO rotational lines are all optically thick, with $\tau\geq10$ at $\Sigma_\mathrm{SFR}>1$~M$_\odot$~yr$^{-1}$ kpc$^{-2}$ \citep{Narayanan2014}.

As the results of PDR modelling depend directly on the reconstructed source-plane surface brightness distribution for each tracer, the robustness of the inferred PDR properties against potential bias in line ratios needs to be assessed. In Appendix~\ref{sec:appendix_C}, we investigate the effect of an extreme (order unity) bias in the measured surface brightness ratios for all the tracers considered; note that this is much larger than the surface brightness bias due to lens modelling for even the lowest-S/N tracer considered. We find that the inferred $G$, $n$(H) values shift by $\leq0.5$~dex even in such an extreme scenario, confirming that the PDR results are robust to (reasonable) bias in measured surface brightness ratios.

Finally, we consider the CMB temperature effects at $z=3$ on the observed line luminosities. Adopting the \citet{daCunha2013} models (for n(H)$\sim10^4$~cm$^{-3}$, see results below), the CO line luminosity is biased by $\leq$10~per cent even for the most affected line (CO(3--2)), below the flux calibration uncertainty. We therefore consider the CMB effect on the inferred PDR models to be negligible.

\subsection{Global PDR model}
\label{sec:pdr_global}

\begin{figure*}
\begin{centering}
\includegraphics[width=15cm, clip=true]{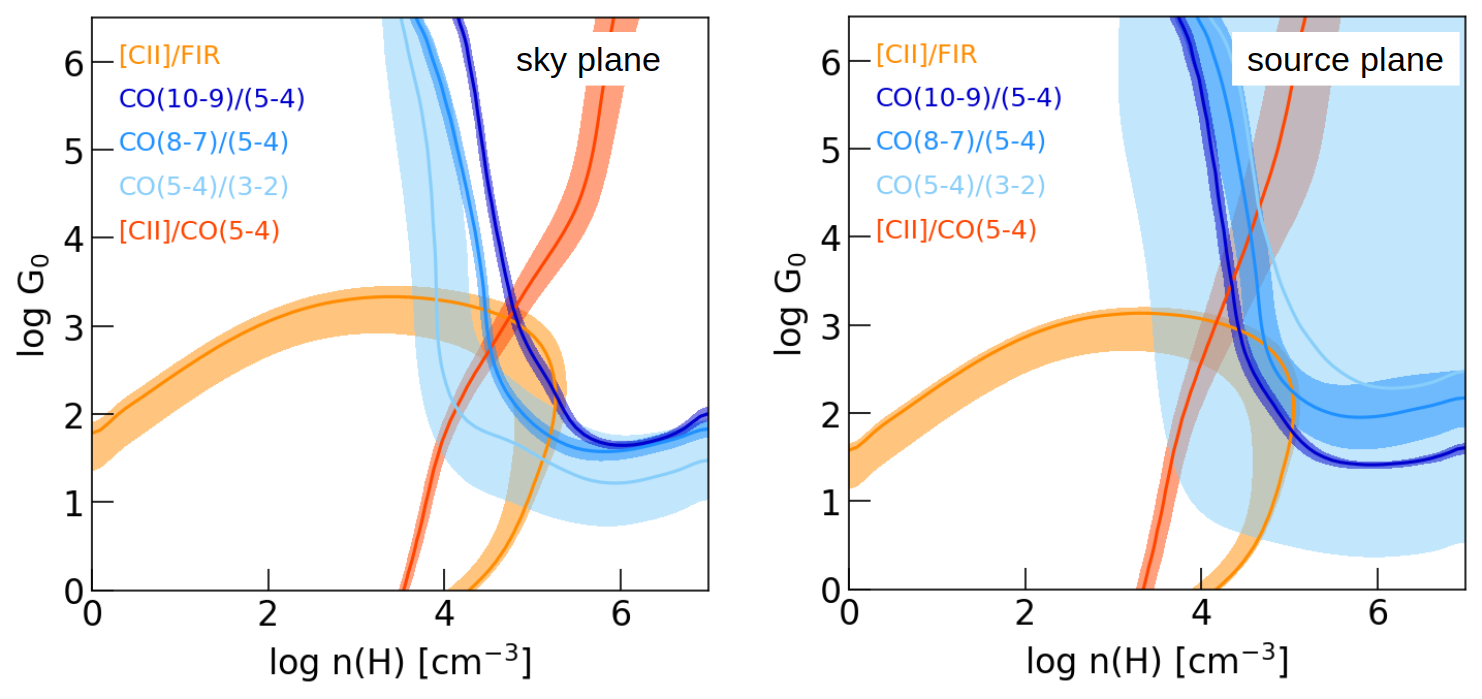}

\caption{Global PDR models for SDP.81, using spatially-integrated source-plane luminosities in the sky plane (\textit{right}) and source plane (\textit{left}), with 1$\sigma$ confidence regions denoted by the colour bands. The differential magnification effect is relatively limited due to the extended nature of all tracers; the best-fitting models differ by $\sim$0.1~dex in $G$ and $n$. \label{fig:pdr_models_global}}
\end{centering}
\end{figure*}

We first apply the PDR modelling to the spatially-integrated source-plane line and continuum luminosities. Fig.~\ref{fig:pdr_models_global} shows the $G$ and $n$ values traces by individual line ratios\footnote{Due to the shape of the [\ion{C}{ii}]/FIR isocontours in the $G_0$/$n$(H) plane, a pure $\chi^2$ optimization might yield very low $G_0$ values for the pixels in the densest regions. This would imply extremely dense regions without any star-formation; we discount these solutions as unphysical. The same issue has been encountered by \citet{Diaz2017} in the analysis of nearby ULIRGs; they likewise discard the low $G_0$/$n$ solutions.}. The maximum a posteriori model gives $G = 10^{3.1\pm0.1}~G_0$, $n = 10^{4.9\pm0.1}$~cm$^{-3}$, with a reduced $\chi^2=0.96$. The derived values are largely insensitive to [\ion{C}{ii}] optical depth: for an optically thick [\ion{C}{ii}] emission, the inferred $G$, $n$ values change by +0.5~dex and -0.1~dex, respectively. Increasing the fraction of [\ion{C}{ii}] emission originating in PDR regions from 80 to 100~per cent results in $G$, $n$ changing by $\leq$0.1~dex.

Compared to the \citet{Valtchanov2011} global PDR model for SDP.81 (inferred using the same {\sc PDRToolbox} models based on [\ion{O}{i}], [\ion{C}{ii}] and CO(1--0) lines and FIR continuum), our $G$, $n$ estimates are $\sim$1~dex higher. This is mainly due to two factors; (1) the lower [\ion{C}{ii}] line luminosity measured by ALMA and the re-reduced \textit{Herschel} spectroscopy (Section~\ref{sec:imaging+modelling}); (2) we are considering CO $J_\mathrm{upp}\geq3$lines, which trace denser molecular gas that CO(1--0) in \citet{Valtchanov2011} (see Section~5.3 for a detailed discussion of systematics).

To assess the impact of differential magnification on the global PDR model, we apply the same analysis to the sky-plane (i.e., not de-lensed) line and FIR luminosities (Fig.~\ref{fig:pdr_models_global}). The best fitting $G$, $n$ sky-plane model is only marginally ($\sim0.1$~dex) offset from the source-plane one. This is due to the large spatial extent (compared to the beam size) of all tracers considered; this significantly reduces the differential magnification bias.

\subsection{Resolved PDR model}
\label{sec:pdr_resolved}

We now perform the resolved PDR modelling, using the line ratios and upper limits for each 200-pc pixel. For this part of the analysis, we adopt conservative uncertainties estimates. The total error budget consists of:
\begin{itemize}
\item Flux calibration uncertainty, assumed to be 10~per cent in each Band.
\item Pixel-by-pixel surface brightness uncertainty from the lens modelling for a fixed source regularization parameter $\lambda_S$, estimated using the scatter in source realizations in Section~\ref{sec:imaging+modelling}.
\item Source-averaged surface-brightness uncertainty estimated from reconstructions of \emph{mock} sources (Appendix~\ref{sec:appendix_B}), which account for potential bias in the maximum a posteriori $\lambda_S$. These vary between 6~per cent (FIR continuum) to 30~per cent for CO(10--9) and 65~per cent for CO(3--2) (Table~\ref{tab:app_b}). For CO(3--2) and CO(10--9), this error dominates the error budget.
\end{itemize}

\begin{figure*}
\begin{centering}
\includegraphics[width=17.5cm, clip=true]{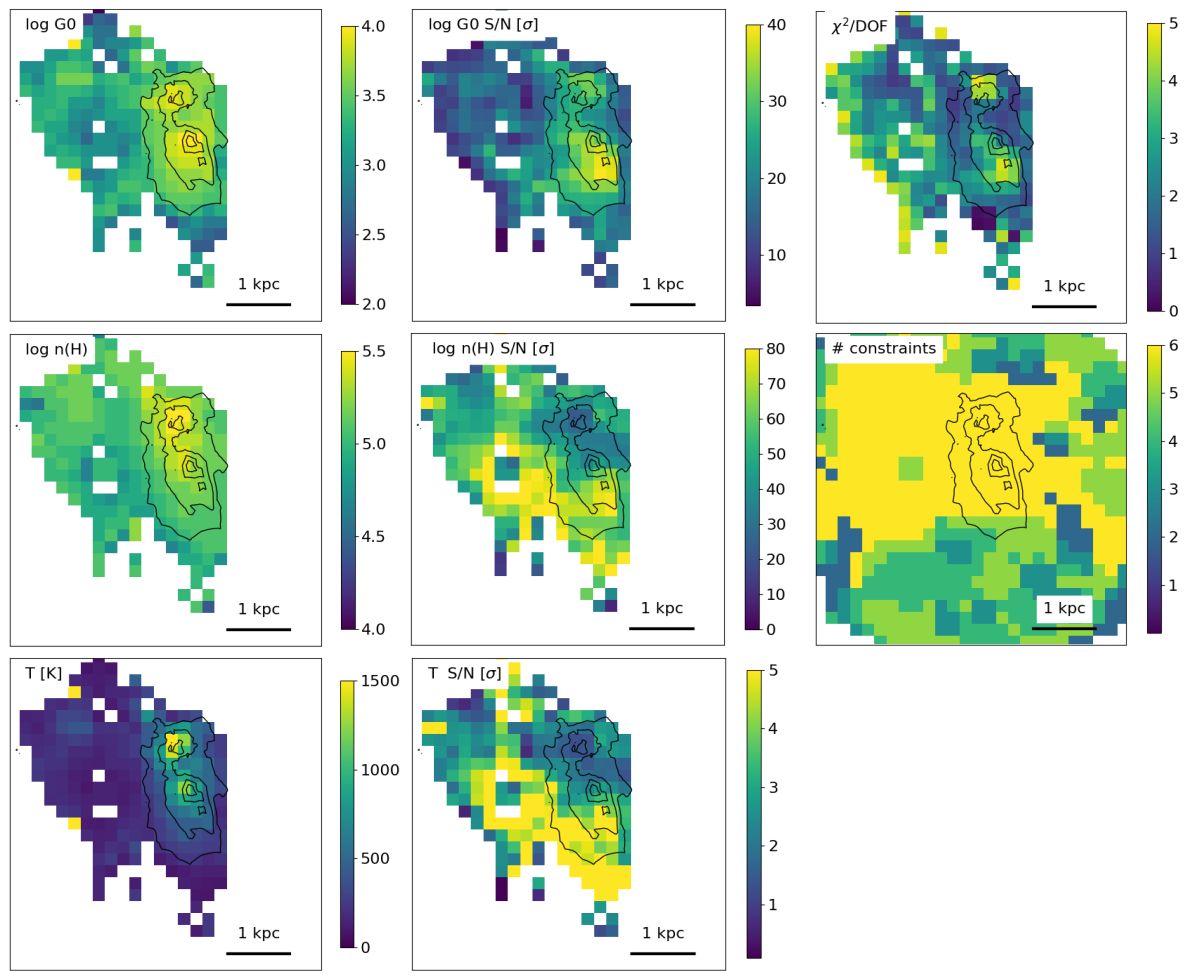}\\

\caption{\textit{Left column}: FUV field strength $G_0$, density $n$(H) and PDR surface temperature $T_\mathrm{PDR}$ inferred from a comparison with the {\sc PDRToolbox} models \citep{Kaufman1999,Kaufman2006,Pound2008}, with the corresponding 1$\sigma$ uncertainties (\textit{middle column}), reduced $\chi^2$ per pixel and the number of line ratios used to constrain the fit for each pixel (\textit{right column}). The pixel size is 200$\times$200~pc. In the FIR-bright region (as denoted by black contours), $G\simeq10^{3.5}-10^{4.0} G_0$, $n$(H)$\simeq10^{4.5}-10^{5.0}$ cm$^{-3}$. The FIR-bright regions of the source shows $T_\mathrm{PDR}$ much higher than the [\ion{C}{ii}] transition temperature (92~K). The pixels for which the inferred $G_0$ and $n$ have S/N<3 and/or with reduced $\chi^2\geq5$ have been masked for clarity.\label{fig:G0_nH}}
\end{centering}
\end{figure*}

During the fitting procedure, we discard the solutions with $G_0/n$(H)$\leq10^{-3}$, which would imply an unphysically large source (c.f.~\citealt{Diaz2017, Wardlow2017}).
Fig.~\ref{fig:G0_nH} shows the $G_0$ and $n$(H) values inferred for individual 200-pc regions in SDP.81, given the surface brightness distributions from Fig.~\ref{fig:reconstructions}. In the central, FIR-bright region, we find strong FUV fields with $G=10^{3.5} - 10^{4.0}~G_0$, and high density $n$(H)=$10^{5.0}-10^{5.5}$ cm$^{-3}$, with a typical uncertainty of $\sim0.2$~dex. These agree well with the global PDR model. We find a significant ($>3\sigma$) gradient in $G$ in the East-West direction: $G$ increases from $10^{3.0\pm0.2}$ to $10^{4.0\pm0.1}~G_0$. While the star-burst region might exhibit increased gas density, we do not find a significant spatial variation in $n$(H). The most direct interpretation is that the star-formation is constrained to the central, disc-like part of the gas reservoir.

The PDR surface temperature $T_\mathrm{PDR}$ varies from about 150~K in the outskirts to 1500~K in the FIR-bright clumps. However, with $T_\mathrm{PDR}$ S/N $\leq$5 across the source, we can not robustly measure the spatial variation of $T_\mathrm{PDR}$. The full implications of the high $T_\mathrm{PDR}$ and its local variations for the [\ion{C}{ii}]/FIR deficit are discussed in Section~\ref{subsec:CII_FIR}.

\begin{figure}
\begin{centering}
\includegraphics[width=8.5cm, clip=true]{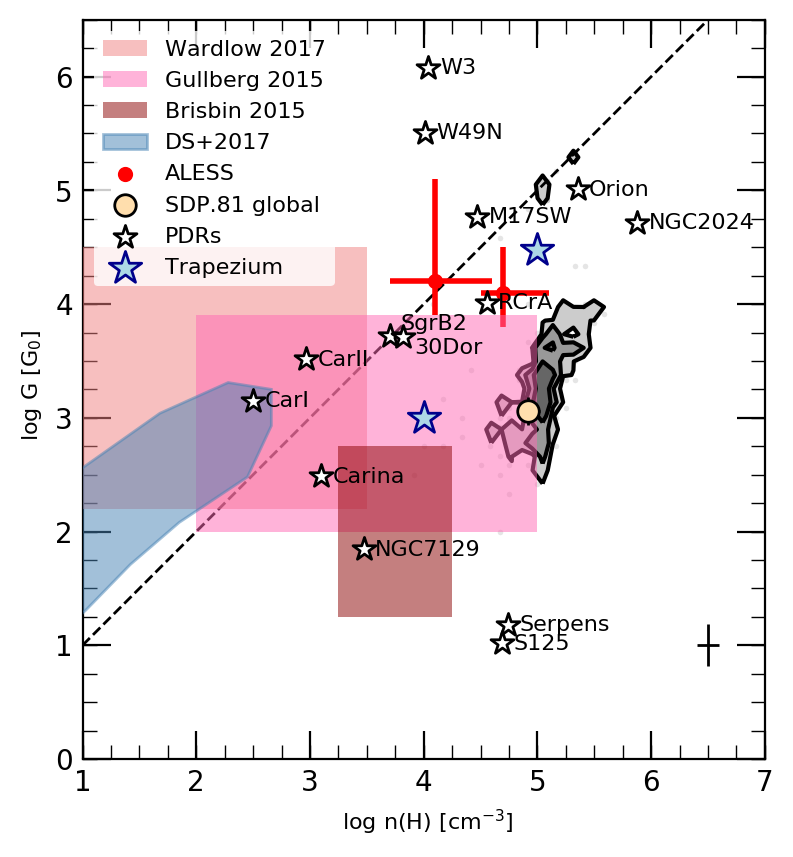}\\

\caption{Resolved $G$, $n$ distribution in SDP.81 (grey, contours drawn at the 25th, 50th and 75th percentile), compared to the global model (Section~\ref{sec:pdr_global}), unresolved values inferred for DSFGs samples of \citet{Brisbin2015}, \citet{Gullberg2015} and \citet{Wardlow2017}, and central regions of two ALESS $z\sim3$ DSFGs \citep{Rybak2019}, $z\sim0$ ULIRGs \citep{Diaz2017}, nearby star-forming regions (adopted from \citealt{Oberst2011}) and resolved measurements in the vicinity of the Trapezium cluster in Orion \citep{Goicoechea2015}. Black errorbars indicate the median 1$\sigma$ uncertainties for the SDP.81 data. The diagonal dashed line indicates the boundary of the region where the dust drift velocity exceeds the average turbulent velocity assumed by the {\sc PDRToolbox} models \citep{Kaufman1999}; above these line, the PDR model becomes internally inconsistent. \label{fig:G0_nH_comparison}}
\end{centering}
\end{figure}

How do these values compare to measurements for other star-forming systems? 
Fig.~\ref{fig:G0_nH_comparison} compares the resolved $G$, $n$ value in SDP.81 to unresolved PDR models of $z=2-5$ DSFGs from \citet{Gullberg2015} and stacked DSFGs from \citet{Wardlow2017}. Compared to both samples, SDP.81 shows a similar FUV field strength, with density $\geq1$~dex above the \citet{Wardlow2017} range, and comparable to the densest \citet{Gullberg2015} sources. Using resolved [\ion{C}{ii}] and CO(3--2) and FIR ALMA imaging, \citet{Rybak2019} inferred $G\sim10^4~G_0$, $n\sim10^4-10^5$~cm$^{-3}$ in the central regions of two $z=2.94$~DSFGs from the ALESS sample; SDP.81 shows slightly higher $G$ and lower $n$(H).

Compared to the PDR properties of the present-day ULIRGs (GOALS sample, \citealt{Diaz2017}) -- inferred from the [\ion{O}{i}] 63-$\mu$m and [\ion{C}{ii}] lines and FIR continuum -- both the global and resolved PDR properties of SDP.81 show density almost $\sim$2~dex higher, with somewhat ($\sim0.5$~dex) stronger FUV fields.

Finally, we consider star-forming regions in the Milky Way and the Large Magellanic Cloud\footnote{Note that the $G$, $n$(H) estimates for local star-forming regions are predominantly inferred from the FIR continuum and the [\ion{C}{ii}], [\ion{O}{i}] 63/145-$\mu$m lines, rather than the CO lines used in this work, and using a variety of modelling approaches. The sizes of local star-forming regions shown here range between $\sim$0.1~pc to $\sim$100~pc.} from the \citet{Oberst2011} compilation. The resolved PDR properties of SDP.81 show densities comparable to large molecular clouds such as Sgr~B2 and 30~Dor, albeit with FUV fields $\sim1$~dex stronger; these might indicate a closer distance between the stars providing FUV radiation and the PDR surfaces, compared to 11--80~pc in 30~Dor (\citealt{Chevance2016}). Indeed, close associations of the O/B stars and gas clouds have been invoked by \citet{Herrera2018b} to explain the [\ion{C}{ii}]/FIR deficit in $z\sim0$ star-forming galaxies and ULIRGs.

On sub-pc scales, we compare the inferred PDR properties with the resolved ($\sim0.05$~pc) study of Orion by \citet{Goicoechea2015}, who used [\ion{C}{ii}], FIR continuum and CO(2--1) observations to infer the PDR conditions in the OMC~1. Using the \citet{Stacey2010} diagnostic diagram (based on \citealt{Kaufman1999, Kaufman2006} PDR models used in our analysis), \citet{Goicoechea2015} obtain $G\simeq3\times10^4~G_0$, $n\geq10^5$~cm$^{-3}$ for the region closest ($R\leq$0.1~kpc) to the Trapezium cluster, decreasing down to $G\simeq10^3~G_0$, $n\simeq10^4$~cm$^{-3}$. These are, respectively, on the upper and lower end of the conditions seen in the FIR-bright region of SDP.81; however, in SDP.81, these describe physical conditions averaged over 200-pc scales (which likely include a number of star-forming regions and the voids between them), underlining the extreme nature of the star-forming regions in DSFGs.

\subsubsection{Caveats and limitations}

Our PDR analysis - both resolved and unresolved - is based on semi-infinite slab, uniform-density PDR models, which are compared with the FIR continuum (160-330~$\mu$m rest-frame), [\ion{C}{ii}] and CO ($J_\mathrm{upp}\geq3$) observations. Our choice of the modelling suite - the {PDRToolbox} models - was motivated by the satisfactory performance of the models given the lines studied here and corresponding the uncertainties, and for consistency with earlier, source-averaged studies of ISM properties in DSFGs. The {\sc PDRToolbox} models reproduce the observed line ratios well (reduced $\chi^2\leq5$), at least given the spatial resolution, diagnostics considered and S/N of the data at hand. Here, we summarize the limitations of this approach:
\begin{itemize}

\item Large surface-brightness uncertainties for CO(3--2) and CO(10--9). As outlined at the beginning of this Section, we adopt conservative surface-brightness uncertainty estimates to avoid over-interpreting the data. For the CO(3--2) and CO(10--9) lines, these are dominated by the error on source reconstruction due to their low S/N. Namely, excluding the uncertainties from Appendix~\ref{sec:appendix_B} from the error budget decreases the quality of the best-fitting models in the FIR-bright clumps(reduced $\chi^2=10-25$). This suggests that an additional heating mechanism or a multi-phase ISM model would be more appropriate in high-$\Sigma_\mathrm{SFR}$ clumps. However, as shown in Appendix~\ref{sec:appendix_B}, the S/N of the CO(3--2) and (10--9) data at hand is currently insufficient to fully constrain the relative importance of different heating mechanisms.

 \item Geometry effects. Our semi-infinite, uniform-density slab models (even after adjusting for multi-side illumination) are likely too simplistic given the complex, clumpy star-forming environment expected in SDP.81, where strong gradients in ISM density and composition along the line-of-sight might be expected. These can be properly accounted for by 3D radiative transfer codes that account for the full spatial stratification of the (multi-phase) ISM, and a realistic distribution of radiation sources. 
 
 \item Isochoric models. The self-gravity and external pressure on the molecular clouds will result in a density stratification, which is not included in our models. The effects of assumed cloud density profiles were recently studied by \citet{Popping2019}, who found that different assumed density profiles result in [\ion{C}{ii}] and CO ($J_\mathrm{upp}\leq5$) luminosities changing by up to $\sim$0.5~dex. This effect might be even more dramatic for the CO(8--7) and CO(10--9) lines. Alternatively, isobaric models - e.g., as in a recent application of the \citep{LePetit2006} code to the resolved observations of 30~Doradus by \citealt{Chevance2016} - might provide a better description of molecular clouds in intensely star-forming environment. However, \citet{Chevance2016} do not find significant differences between gas properies inferred using the isochoric and isobaric assumptions.
 
 \item Reliance on mid- to high-$J$ CO lines. Our PDR modelling is largely driven by the CO(5--4) and CO(8--7) lines, however, the predictions for the emergent line intensities can vary significantly between different PDR codes. This is because the high-$J$ CO emission is confined to the outer cloud layer; therefore, differences in CO abundance will result in a different predictions (e.g., \citealt{Rollig2007}). Therefore, the inferred gas properties might depend strongly on the underlying PDR model.

 \item Limited sensitivity to cold, low-density gas. By including only $J_\mathrm{upp}\geq3$ CO lines (critical density $n_\mathrm{crit}\geq8\times10^3$~cm$^{-3}$, we effectively discount the extended cold, low-density ($n\leq10^3$~cm$^{-3}$) gas phase traced by CO(1--0) and CO(2--1) lines. Indeed, studies of the CO ladders including the $J_\mathrm{upp}=1-2$ lines have found significant cold gas components both in $z\sim0$ ULIRGs (e.g.,~\citealt{Kamenetzky2014, Greve2014}) and high-redshift DSFGs (e.g., \citealt{Bothwell2013, Yang2017}). While the cold gas component can contribute to the CO(3--2) and CO(5--4) luminosity, the \citet{Yang2017} source-averaged CO SLED model for SDP.81 - which includes both a cold and a warm component - shows the cold-gas contribution to the CO(3--2) and CO(5--4) luminosities to be $\leq$30 and $<$10~per cent, respectively. Therefore, our results should not be significantly affected by the cold gas component (Appendix~\ref{sec:appendix_C}).
 
 \item Mechanical and cosmic/X-ray heating. In the high-$\Sigma_\mathrm{SFR}$ regime, additional mechanism not included in the {\sc PDRToolbox} models - e.g., mechanical heating by shocks, X-ray radiation and cosmic rays - can contribute significantly to the total gas heating. Although the line ratios in SDP.81 are well-reproduce by models that include only far-UV heating by stars; some of the key diagnostics of e.g., mechanical heating, are not available for SDP.81 (Section 4.2.1). Future observations of additional tracers (e.g., high-$J$ CO lines, mid-$J$ HCN/HCO$^+$ lines, CO isotopologues) will be necessary to properly assess the relative importance of different heating mechanisms.
 
\end{itemize}

\subsection{What drives the [CII]/FIR deficit in SDP.81?}
\label{subsec:cii_fir_origin}

With the inferred $G$, $n$ and $T_\mathrm{PDR}$ maps in hand, we now address the origin of the [\ion{C}{ii}]/FIR deficit in SDP.81.

Recently, \citet{Rybak2019} used resolved (0.15--0.5~arcsec) observations of [\ion{C}{ii}], CO(3--2) and FIR continuum emission in two unlensed $z\sim3$ DSFGs to show that the [\ion{C}{ii}]/FIR deficit in their central regions is due to thermal saturation. We apply the analysis of \citet{Rybak2019} to the 200-pc resolution maps of SDP.81. Namely, we consider the following mechanisms: (1) thermal saturation of the [\ion{C}{ii}] fine-structure levels; (2) reduced gas heating due to positive grain charging; (3) AGN contribution.

\subsubsection{Reduced photoelectric heating}

We calculate the photoelectric heating rate per hydrogen atom following \citet{Wolfire2003}:

\begin{multline}
\Gamma_\mathrm{PE} = \frac{1.1\times10^{-25}G'Z_d}{1+3.2\times10^{-2}\Big[G'\big(T_\mathrm{gas}/100~K\big)^{1/2} n_e^{-1}\phi_\mathrm{PAH}\Big]^{0.73}} \\ \mathrm{erg \, s^{-1}},
\label{eq:photoel}
\end{multline}

where $G'=1.7 \times G_0$, $Z_d$ is the dust-to-gas ratio (normalized to the Galactic value), $n_e$ the electron density and $\phi_\mathrm{PAH}$ a factor associated with the PAH molecules. Assuming $n_e=1.1\times10^{-4}\,n$ and $\phi_\mathrm{PAH}=0.5$, we calculate $\Gamma_\mathrm{PE}$ for each resolution 200$\times$200~pc resolution element. The $\Gamma_\mathrm{PE}$ varies by only $\sim$50~per cent across the source; this variation is insufficient to explain the $\sim$1.5~dex variation in the observed [\ion{C}{ii}]/FIR ratio.

\subsubsection{AGN effects}
At sub-kpc scales, a strong [\ion{C}{ii}]/FIR deficit can be driven by AGNs, by simultaneously suppressing the [\ion{C}{ii}] emission by X-ray-driven ionization of C$^+$ to C$^{2+}$ etc., and increasing the FIR luminosity due to extra FUV flux from the accretion disk. Resolved [\ion{C}{ii}]/FIR studies of local active galaxies \citep{Smith2017, Herrera2018b} have revealed a strong [\ion{C}{ii}]/FIR drop within $\sim500$-pc radius of the AGN in some galaxies. While the AGN influence will be largely negligible for kpc-scale deficits (c.f., \citealt{Rybak2019}), the AGN sphere of influence is comparable to the 200-pc resolution of our data.
XMM observations of the SDP.81 field detect 0.5--8~keV flux of $(49\pm7)\times10^{-15}$ erg s$^{-1}$ cm$^{-2}$, with a hardness ratio of -0.4 at the position of SDP.81 \citep{Ranalli2015}. However, it is unclear whether this flux is associated with the lensed DSFG, or the AGN in the $z=0.299$ lens, which is detected in radio and mm-wave imaging. Assuming the X-ray flux is all due to a hypothetical AGN at the position of the northern FIR-bright clump (lensing magnification $\mu\simeq20$), we find a source-plane X-ray luminosity $L_\mathrm{0.5-8.0~keV}\simeq2\times10^{44}$~erg s$^{-1}$ (uncorrected for obscuration). This is similar to X-ray luminosities observed in DSFGs from the ALESS sample ($L_\mathrm{0.5-8.0~keV}=23\times10^{42}-3\times10^{44}$~erg s$^{-1}$, \citealt{Wang2013})\footnote{Based on the \citet{Wang2004} models, the observed hardness ratio in SDP.81 is compatible with both a low-obscuration ($N_\mathrm{H}\simeq10^{21.5}$ cm$^{-3}$, $A_\mathrm{V}\simeq3$) $z=0.299$ solution, and a highly obscured ($N_\mathrm{H}=10^{23}$ cm$^{-3}$, $A_\mathrm{V}\simeq5$) $z=3.04$ solution. We derive the corresponding $A_\mathrm{V}$ factors using the \citet{Guver2009} $N_\mathrm{H}-A_\mathrm{V}$ relation}. Even assuming the highest column density for the ALESS sources from \citet{Wang2013}, ($N_\mathrm{H}\simeq5\times10^{24}$~cm$^{-3}$), a comparison with carbon ionization model in the vicinity of X-ray bright AGNs of \citet{Langer2015} shows that the X-ray flux in SDP.81 is insufficient to drive the [\ion{C}{ii}]/FIR deficit on scales beyond few hundred pc (c.f., \citealt{Rybak2019}). Finally, circumstantial evidence also disfavours the AGN-driven [\ion{C}{ii}]/FIR deficit: (1) a single AGN is hard to reconcile with the two similarly deep [\ion{C}{ii}]/FIR minima separated by $\sim$1~kpc (Fig.~\ref{fig:cii_fir_comparison}; although an ''Arp~220``-like merging configuration would account for this) and (2) although the CO SLED is more excited in the northern part of the FIR-bright region (as evidenced by the higher CO(10--9) surface brightness, Fig.~\ref{fig:reconstructions}), the CO excitation is fully consistent with star-formation, without a need for additional AGN-associated X-ray/cosmic-ray heating. 

\subsubsection{C$^+$ thermal saturation}

At high gas temperature $T_\mathrm{gas}$, the relative occupancy of the upper/lower C$^+$ fine-structure levels becomes thermally saturated \citep{MunozOh2016} at temperatures higher than the [\ion{C}{ii}] transition temperature $T_\mathrm{[CII]}=92$~K. As the C$^+$ requires high UV flux to remain ionized, the C$^+$ (and [\ion{C}{ii}] emission) is constrained to the surface layer ($A_V\leq2$) of the molecular clouds. This allows us to take the PDR surface temperature $T_\mathrm{PDR}$ as indicative of the gas temperature in the [\ion{C}{ii}]-emitting regions. For a dense gas ($n$ larger than [\ion{C}{ii}] critical density) and assuming the [\ion{C}{ii}] line is optically thin, the gas cooling via the [\ion{C}{ii}] line per hydrogen atom scales as:

\begin{equation}
\Lambda_\mathrm{[CII]} = 2.3\times10^{-6} \, k_\mathrm{B} \, T_\mathrm{[CII]} \frac{2}{2+\exp(T_\mathrm{CII}/T_\mathrm{gas})} \frac{\mathrm{C}}{\mathrm{H}},
\label{eq:cii_cooling}
\end{equation}

where $k_\mathrm{B}$ is the Boltzmann constant, and C/H the carbon-to-hydrogen number ratio (see \citealt{MunozOh2016}).

\begin{figure}
\begin{centering}
\includegraphics[width=8.5cm, clip=true]{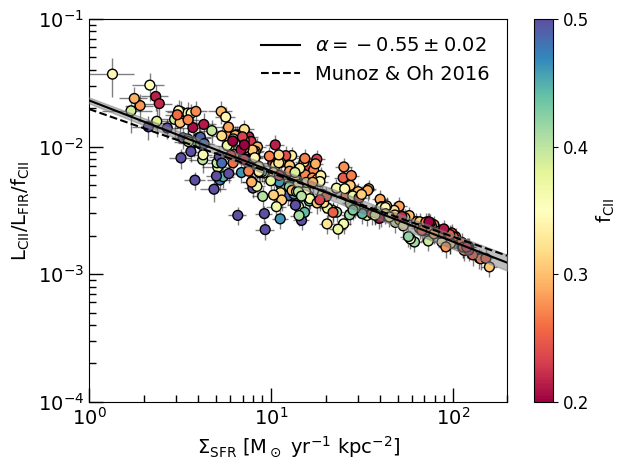}\\

\includegraphics[width=8.5cm, clip=true]{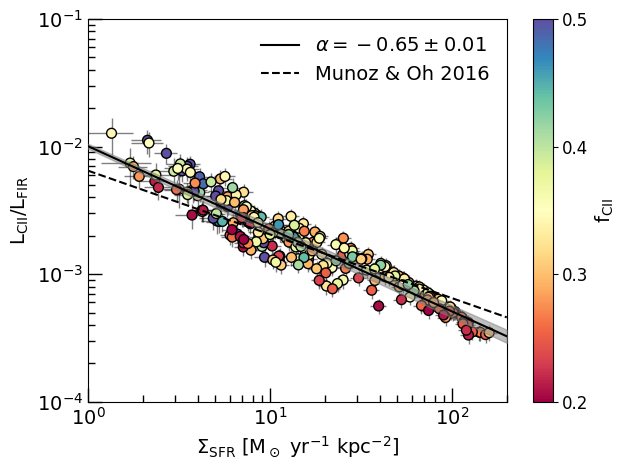}\\

\caption{Comparison of the [\ion{C}{ii}]/FIR ratio in SDP.81, corrected for $f_\mathrm{[CII]}$, the fraction of gas traced by the [\ion{C}{ii}] (upper), and without the correction (lower), with the thermal saturation model of \citet{MunozOh2016} (dashed line). The solid line and the shaded region indicate the best-fitting power-law function to the data with the associated $1\sigma$ confidence region. The $f_\mathrm{[CII]}$ correction significantly reduces the tension between the observed [\ion{C}{ii}]/FIR ratio and the thermal saturation model. \label{fig:cii_fir_slopefit}}
\end{centering}
\end{figure}

In the thermally saturated regime \citep{MunozOh2016}, the [\ion{C}{ii}]/FIR ratio is
\begin{equation}
\frac{L_\mathrm{[CII]}}{L_\mathrm{FIR}} = 2.2\times10^{-3} f_\mathrm{[CII]} \left[\frac{\Sigma_\mathrm{FIR}}{10^{11}~L_\odot~\mathrm{kpc}^{-2}}\right]^{-1/2},
\label{eq:cii_munozoh2016}
\end{equation}

where $f_\mathrm{[CII]}$ is the mass of gas traced by the [\ion{C}{ii}] emission (\citealt{MunozOh2016}):

\begin{equation}
f_\mathrm{[CII]} = \frac{1}{1+3\times10^4 \alpha_\mathrm{CO} L_\mathrm{CO(1-0)}/L_\mathrm{CII}},
\label{eq:f_cii}
\end{equation}

assuming the fraction of gas traced by [\ion{C}{ii}] or doubly-ionized carbon is negligible.
As noted already in Section~\ref{subsec:CII_FIR}, the $L_\mathrm{[CII]}/L_\mathrm{FIR}$ slopes in SDP.81 (this work) and SDP.11 \citep{Lamarche2018} are considerably steeper (approximately -0.7), compared to the -0.5 slope in Equation (\ref{eq:cii_munozoh2016}. However, $f_\mathrm{[CII]}$ is likely varying across the source. We quantify this directly via Equation~(\ref{eq:f_cii}), estimating $L_\mathrm{CO(1-0)}$ from the resolved $L_\mathrm{CO(5-4)}$ maps\footnote{Although using the CO(3--2) emission would give us a more direct probe of CO(1--0), its S/N is insufficient to trace the small-scale structure seen in the [\ion{C}{ii}]/FIR maps.} by using the \citet{Bothwell2013} conversion factor and $\alpha_\mathrm{CO}$ from \citet{CR18} (see Section~\ref{subsec:t_dep}). To ensure a robust $f_\mathrm{[CII}$ estimates and to eliminate the effect of potentially spurious artifacts, we consider the pixels with CO(5--4) S/N$\geq3$, and within the region denoted in Figure~\ref{fig:G0_nH}. Calculating the $f_\mathrm{[CII]}$ correction for each 200$\times$200~pc pixel, we obtain a mean $f_\mathrm{[CII]}=0.32\pm16$. After correcting the resolved [\ion{C}{ii}]/FIR deficit for $f_\mathrm{[CII]}$ (Fig.~\ref{fig:cii_fir_slopefit}, upper panel), we obtain a best-fitting power-law slope of $=-0.55\pm0.02$, in line with the \citet{MunozOh2016} prediction. For comparison, fitting the uncorrected [\ion{C}{ii}]/FIR data over the same region yields $\alpha=0.65\pm0.01$, in a significant tension with the \citet{MunozOh2016} model.

Given the high PDR surface temperature inferred from PDR modelling (Section~\ref{sec:pdr_models}), the good match between the [\ion{C}{ii}] abundance-corrected [\ion{C}{ii}]/FIR slope with the \citet{MunozOh2016} predictions, and the limited effects of the positive grain charging and AGN heating, we consider the thermal saturation of the C$^+$ fine-structure levels to be the most likely cause of the strong [\ion{C}{ii}]/FIR deficit in SDP.81.

\section{Conclusions}
\label{sec:conclusions}

We have presented matched-resolution source-plane reconstructions of the [\ion{C}{ii}], CO (3--2), CO(10--9) and the ALMA Bands 6, 7 and 8 (rest-frame 160-320~$\mu$m) continuum emission in a $z=3.042$ gravitationally lensed, dusty star-forming galaxy SDP.81. Using a visibility-fitting lens modelling code with an adaptive source-plane grid \citep{Rybak2015a}, we achieve a median source-plane resolution of $\sim$200~pc. Combining these with the CO(5--4) and CO(8--7) data from \citet{Rybak2015b}, we use PDR models of \citet{Kaufman1999, Kaufman2006, Pound2008}, to map the physical conditions of the star-forming ISM -- in particular, the FUV field strength, gas density and PDR surface temperature -- on sub-kpc scales. Compared to other studies of the [\ion{C}{ii}]/FIR deficit at high redshift, we are able to leverage the resolved CO and FIR continuum observations to characterize the ISM properties in SDP.81 on sub-kpc scales. Our main results are:

\begin{itemize}

\item The ALMA Band~8 continuum source-plane surface brightness distribution is consistent with the previously-published Band~6 and Band~7 continuum reconstructions (R15a, \citealt{Dye2015, Hezaveh2016}). Fitting for the dust temperature on 200-pc scales, we do not find evidence for a strong dust temperature gradient, and $T_\mathrm{dust}\geq80$~K, in tension with models proposed by \citet{CR18}.

\item The [\ion{C}{ii}] emission is significantly more extended than the FIR continuum and CO($J_\mathrm{upp}\geq5$) emission: $\sim$50~per cent of [\ion{C}{ii}] emission originates outside the FIR-bright region. The [\ion{C}{ii}] maps reveal an extended, low surface-brightness emission extending to $\sim$4~kpc from SDP.81, providing further evidence to SDP.81 being a merger-induced starburst.

\item The CO(3--2) emission is concentrated to the north of the source, with the CO(3--2)/FIR ratio varying rapidly across the source (by $\geq1$~dex). Using the CO(3--2) as a molecular gas tracer, we find an extremely short gas depletion timescale of $t_\mathrm{dep}=10^6-2\times10^7$~Myr, significantly shorter than in most other resolved $t_\mathrm{dep}$ studies in high-redshift DSFGs.

\item The CO SLED in SDP.81 is comparable to that of Class~I ULIRGs from \citet{Rosenberg2015}. Given the data at hand, the CO SLED is consistent with PDR models, without a need for additional energy source (e.g., mechanical heating). Although globally, the CO lines contribute only marginally (few per cent) to the total gas cooling, the CO/[\ion{C}{ii}] ratio can be as high as 25~per cent for the FIR-bright clumps.

\item Comparing the source-integrated (unresolved) de-lensed [\ion{C}{ii}], CO and FIR emission with PDR models of \citet{Kaufman1999, Kaufman2006, Pound2008}, we find FUV field strength $G=10^3~G_0$ and gas density $n$(H)=$10^5$~cm$^{-3}$, significantly higher than the earlier model by \citet{Valtchanov2011}. The source-averaged PDR properties inferred from the de-lensed (source-plane) and observed (sky-plane) luminosities are consistent with each other; the differential lensing does not bias the global PDR model significantly.

\item Modelling the PDR conditions on 200-pc scales, we find the FUV field strength to vary considerably across the source ($G=10^{2.5}-10^{4.0}~G_0$). In contrast, $n$(H) varies only marginally ($10^{5.0}-10^{5.5}$~cm$^{-3}$). Compared to the unresolved models, the resolved PDR model shows $G$ up to 1~dex higher; this is due to the strong variation of [\ion{C}{ii}]/FIR ratio (which determines $G$) on sub-kpc scales, the FIR continuum being much more compact than [\ion{C}{ii}]. The physical conditions averaged over 200-pc scales are comparable to the vicinity of the Trapezium cluster in Orion \citep{Goicoechea2015}, underlying the extreme nature of DSFGs, and highlighting a need for more sophisticated radiative transfer models for interpreting the observed line ratios.

\item The [\ion{C}{ii}]/FIR ratio in SDP.81 varies by $\sim$1.5~dex across the source, with a strong [\ion{C}{ii}]/FIR deficit in the central FIR-bright clumps ($L_\mathrm{[CII]}/L_\mathrm{FIR}=3\times10^{-4}$). Although the [\ion{C}{ii}] surface brightness generally increases towards the FIR-bright part of the source, it drops at the position of the FIR-bright clumps.

\item Given the resolved PDR properties, the most likely cause of the [\ion{C}{ii}]/FIR deficit in SDP.81 is thermal saturation of the C$^+$ fine-level structure due to strong FUV fields and high PDR surface temperature. After correcting the resolved [\ion{C}{ii}]/FIR for the fraction of gas traced by the [\ion{C}{ii}] emission, we find the [\ion{C}{ii}]/FIR trend to be consistent with the thermal saturation model of \citet{MunozOh2016}.

\end{itemize}

We emphasise that the agreement between the observed line ratios and the simplistic PDR models is largely due to significant systematic uncertainties on the CO(3--2) and CO(10--9) source reconstructions. High-sensitivity observations of these two lines, and additional fainter diagnostics (e.g., HCN and HCO$^+$ lines, CO isotopologues) are necessary to discriminate between e.g., different heating mechanism on sub-kpc scales.

This paper presents a first look into the PDR properties of high-redshift, dusty star-forming galaxies on scales of few hundred pc, and a major improvement on the unresolved studies of DSFGs. Our results reveal the complex nature of DSFGs, with strong spatial variations in [\ion{C}{ii}]/FIR ratio, the FUV field strength and the PDR surface temperature. By studying the properties of the photon-dominated regions on 200~pc scales, this work bridges the high-redshift starbursts and local studies of large star-forming regions. 
We make the reconstructed source-plane maps of the CO ladder and [\ion{C}{ii}] line and the ALMA Bands 6, 7 and 8 continuum (Figures~\ref{fig:reconstructions}, \ref{fig:reconstructions_appendix_cii}) publicly available at \texttt{sdp81.strw.leidenuniv.nl}. We hope this legacy dataset will facilitate independent analysis of the conditions in high-redshift DSFGs on sub-kpc scales.

\section*{Acknowledgements}

The authors thank the anonymous referee for their thorough comments and M. Kazadjian and F. Israel for helpful discussions on the radiative transfer modelling, Ana Lopez-Sepulcre for help with ALMA observing setup, and C. Lamarche and K. Litke for sharing their [\ion{C}{ii}]/FIR data. MR thanks MPA~Garching, and S. D. M. White in particular, for the access to computational facilities.

MR and JH acknowledge support of the VIDI research programme with project number 639.042.611, which is (partly) financed by the Netherlands Organisation for Scientific Research (NWO). LG acknowledges support from the Amaldi Research Center funded by the MIUR program "Dipartimento di Eccellenza" (CUP:B81I18001170001). This paper makes use of the following ALMA data: ADS/JAO.ALMA \#2011.0.00016.SV, \#2016.1.00633.S and \#2016.1.01093.S. ALMA is a partnership of ESO (representing its member states), NSF (USA) and NINS (Japan), together with NRC (Canada), MOST and ASIAA (Taiwan), and KASI (Republic of Korea), in cooperation with the Republic of Chile. The Joint ALMA Observatory is operated by ESO, AUI/NRAO and NAOJ.
Fig.~\ref{fig:G0_nH_comparison} utilizes the {\sc corner} package by \citet{Corner}.

\newpage

\appendix

\section{Lensing reconstructions}
\label{sec:appendix_A}

\begin{figure*}

\includegraphics[width=18.0 cm, clip=true]{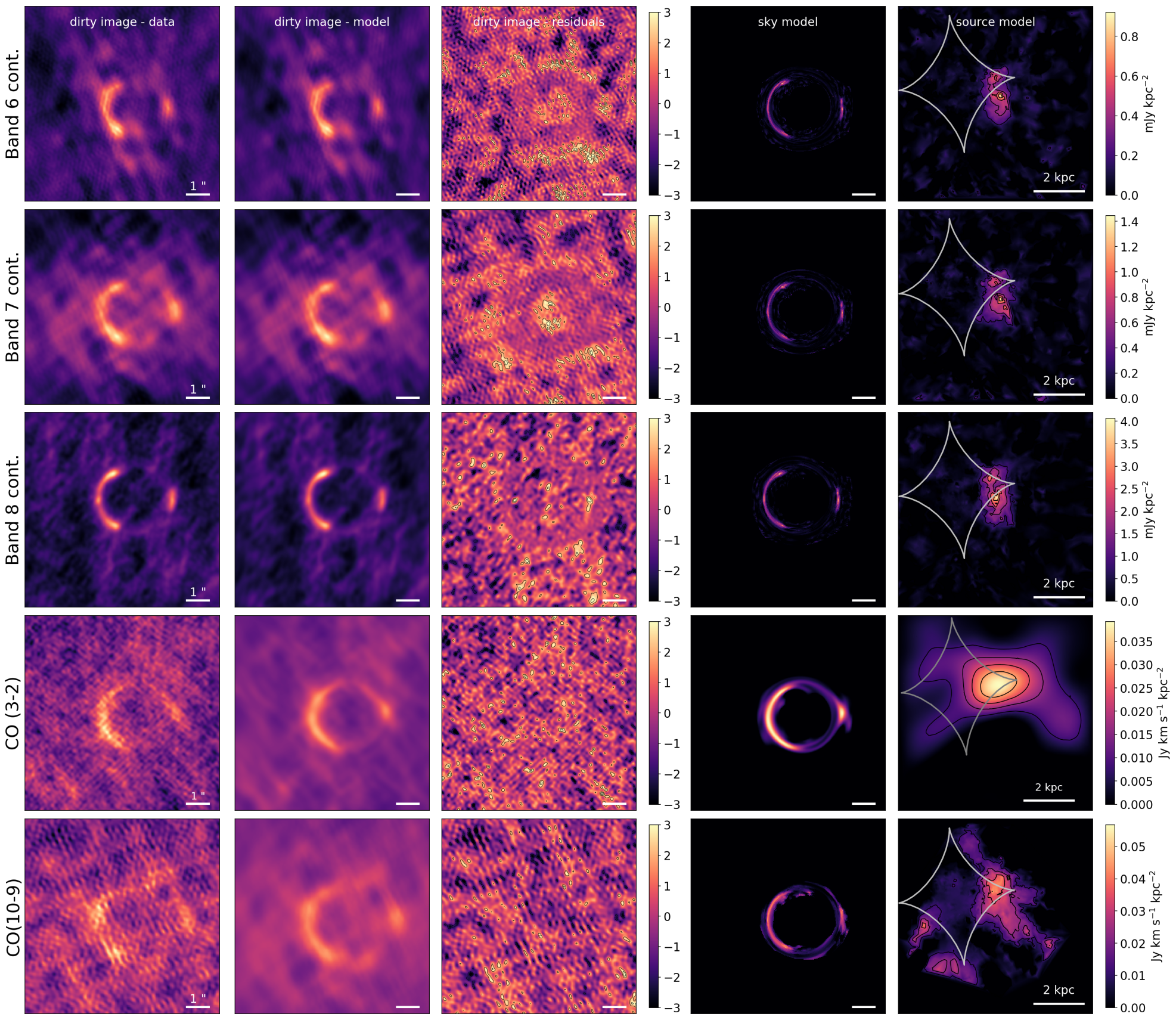}

\caption{Lens modelling results: maximum a posteriori models for (top to bottom): the Bands 6, 7 and 8 continuum, the CO(3--2) and the CO(10--9) lines (\textit{lower}). Individual columns are ordered as follows: (1) dirty image of the data; (2) dirty image of the maximum a posteriori (MAP) model (normalized to the peak surface brightness of the data dirty image); (3) dirty-image residuals, with contours starting at $\pm2\sigma_\mathrm{rms}$ and increasing by steps of $1\sigma_\mathrm{rms}$; (4) MAP sky model (normalized to the peak surface brightness); (5) MAP reconstructed source, with contours drawn at 25, 50 and 75~per cent of the surface brightness maximum. Source-plane caustics are marked by the grey line. \label{fig:reconstructions_appendix}}
\end{figure*}

Figures~\ref{fig:reconstructions_appendix} and \ref{fig:reconstructions_appendix_cii} show the lens modelling results for the Band~8 continuum, velocity-integrated CO(3--2) and CO(10--9) lines, and the reconstructions for individual 40~km s$^{-1}$ [\ion{C}{ii}] velocity channels. The dirty images are calculated using natural weighting (i.e., each visibility is weighted by $1/\sigma^2$).

\begin{figure*}

\includegraphics[width=16cm, clip=true]{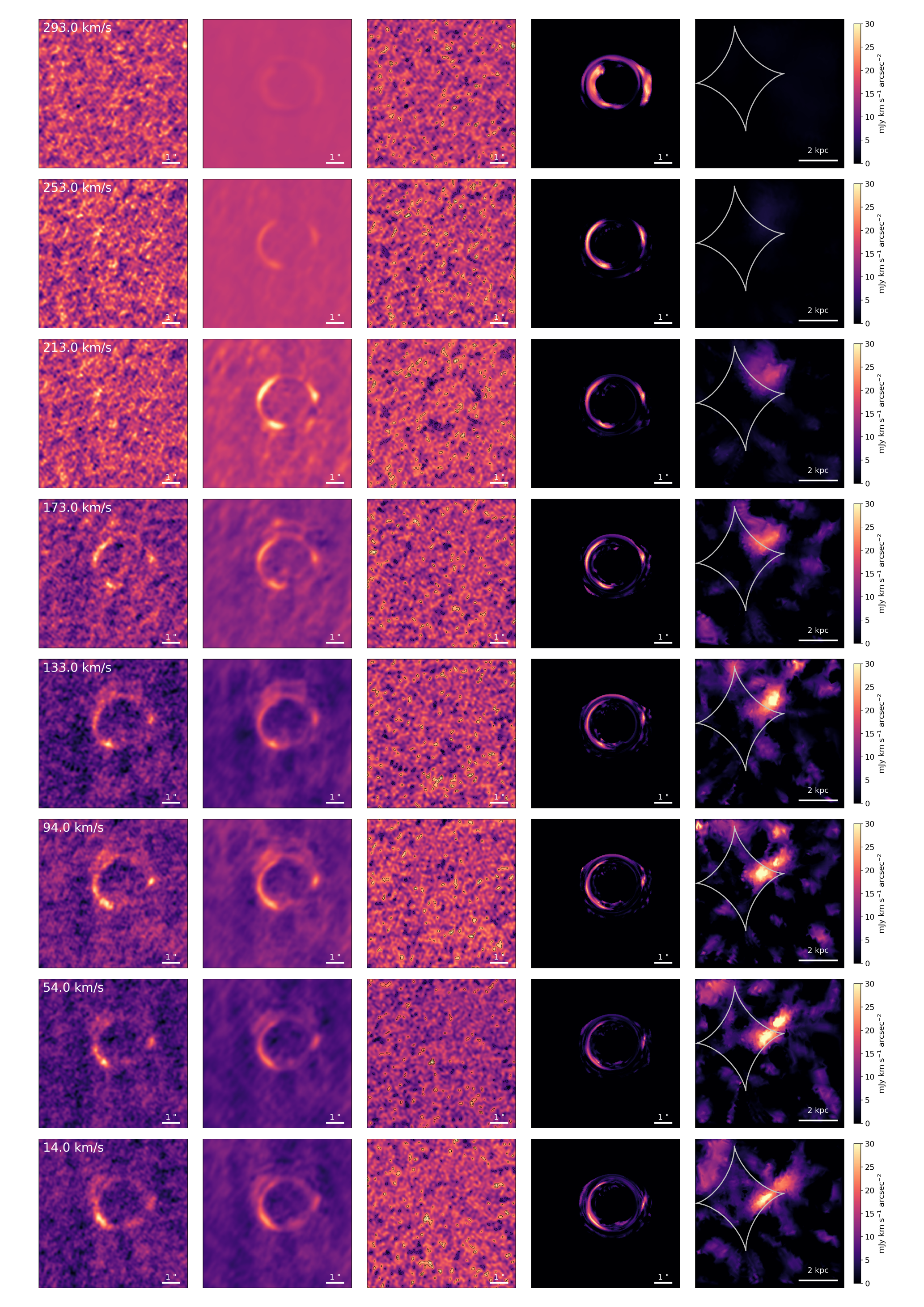}

\caption{[\ion{C}{ii}] line reconstruction, using a channel width of 62.5~MHz ($\sim40$~km s$^{-1}$. Individual panels ordered as in Fig.~\ref{fig:reconstructions_appendix}. Channel velocity is given in LSRK with respect to a systemic redshift $z=3.042$, using the radio definition of velocity. \label{fig:reconstructions_appendix_cii}}
\end{figure*}

\begin{figure*}
\ContinuedFloat
\includegraphics[width=16cm, clip=true]{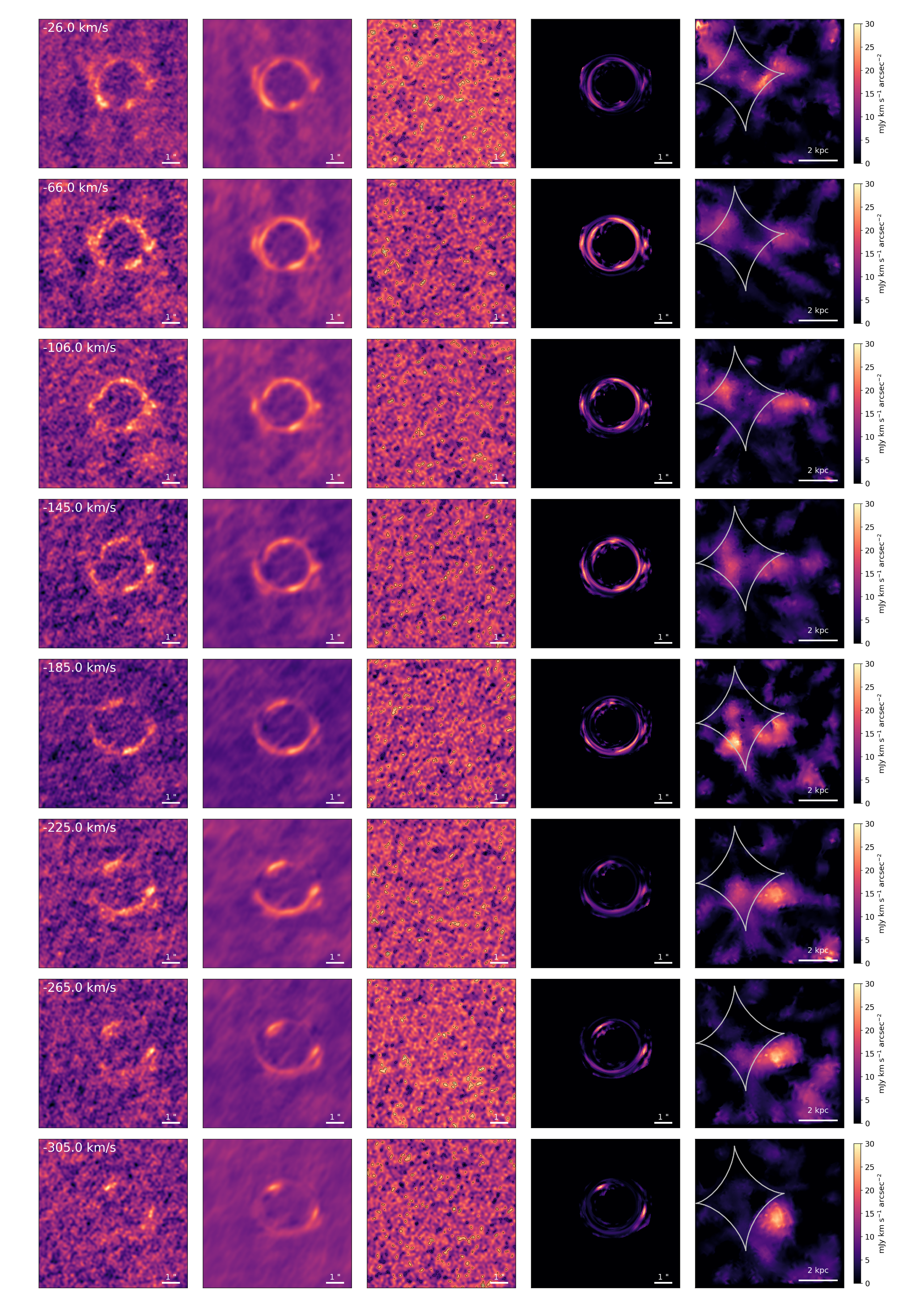}
\caption{Continued.}
\end{figure*}

\begin{figure*}
\ContinuedFloat

\includegraphics[width=16cm, clip=true]{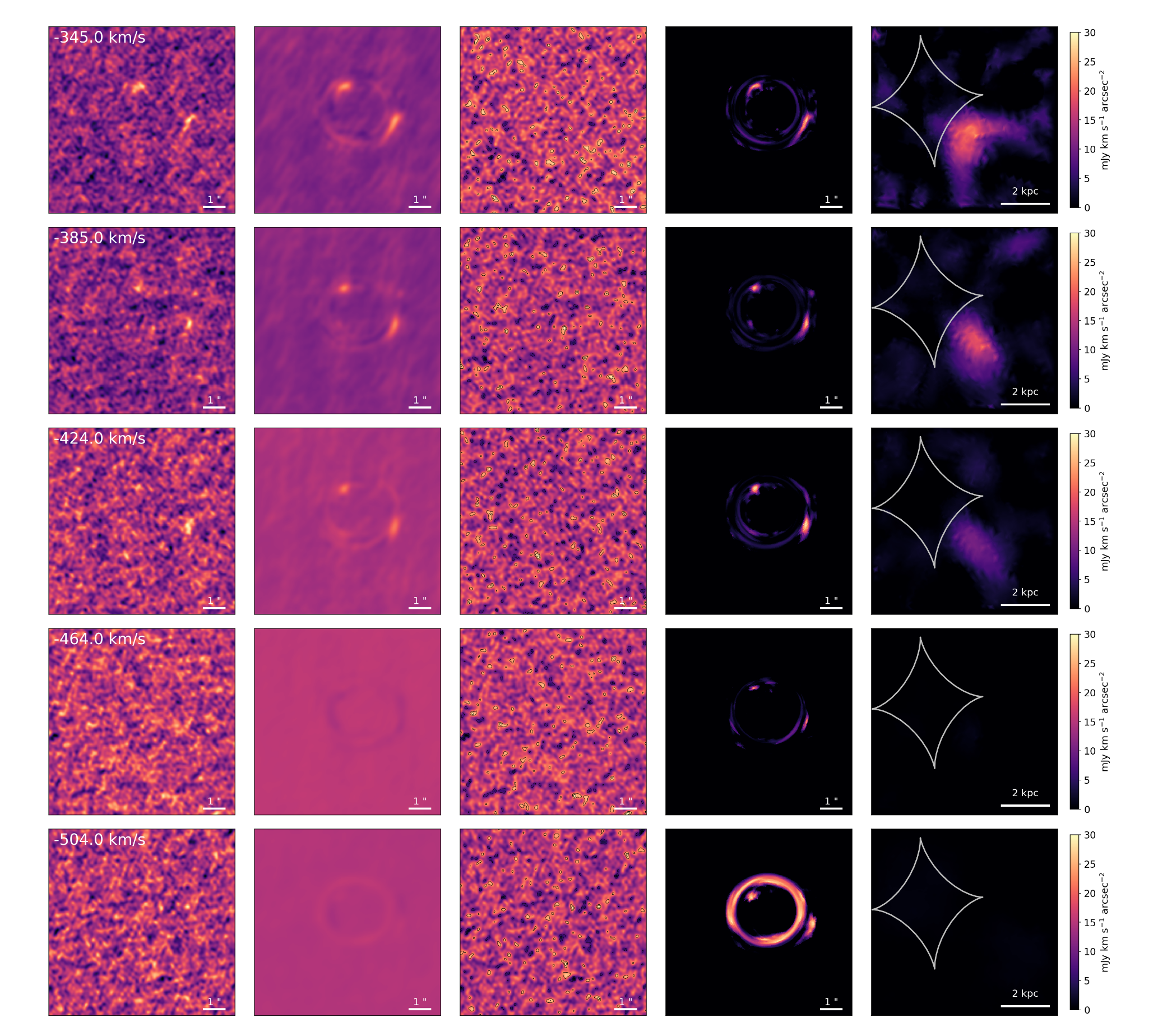}

\caption{Continued.}
\end{figure*}

\section{Estimating the systematic error on the source reconstructions}
\label{sec:appendix_B}

We reconstructed the lensed source using source-plane regularization to impose a spatial correlation in the source-plane and prevent noise-fitting. However, for low-S/N data, the optimization process might yield high levels of regularization, overly smoothing the source. Here, we assess the systematic uncertainty on the reconstructed source by creating mock ALMA observations for each tracer.

\begin{figure*}
\begin{centering}
\includegraphics[width=18cm, clip=true]{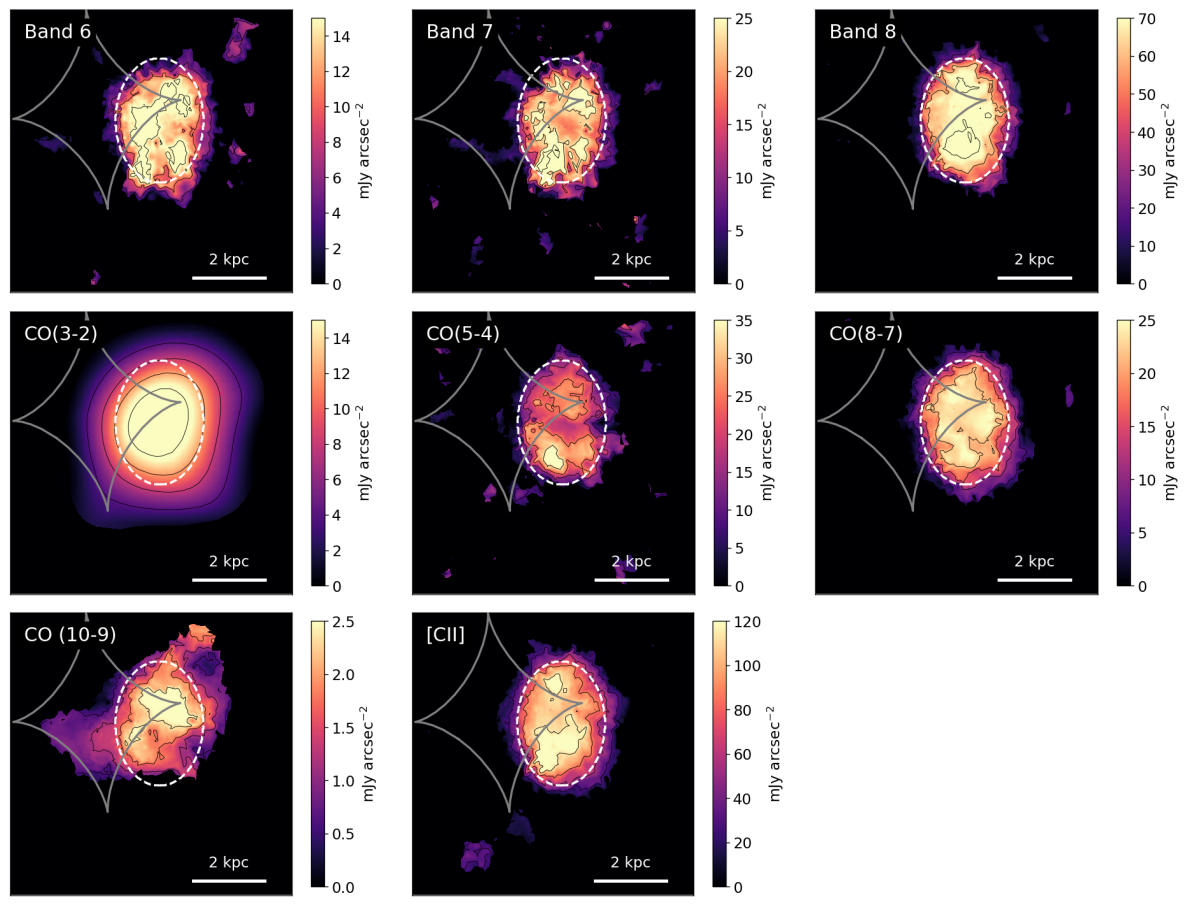}

\caption{\textit{Upper:} Source-plane surface brightness reconstructions for mock ALMA observations of a uniform surface brightness source corresponding to the individual tracers. The input source size is indicated by the white dashed contour. The black contours correspond to 20, 40, 60 and 80~per cent of the peak surface brightness. \label{fig:app_B1}}
\end{centering}
\end{figure*}

\begin{figure*}
\begin{centering}
\includegraphics[width=18cm, clip=true]{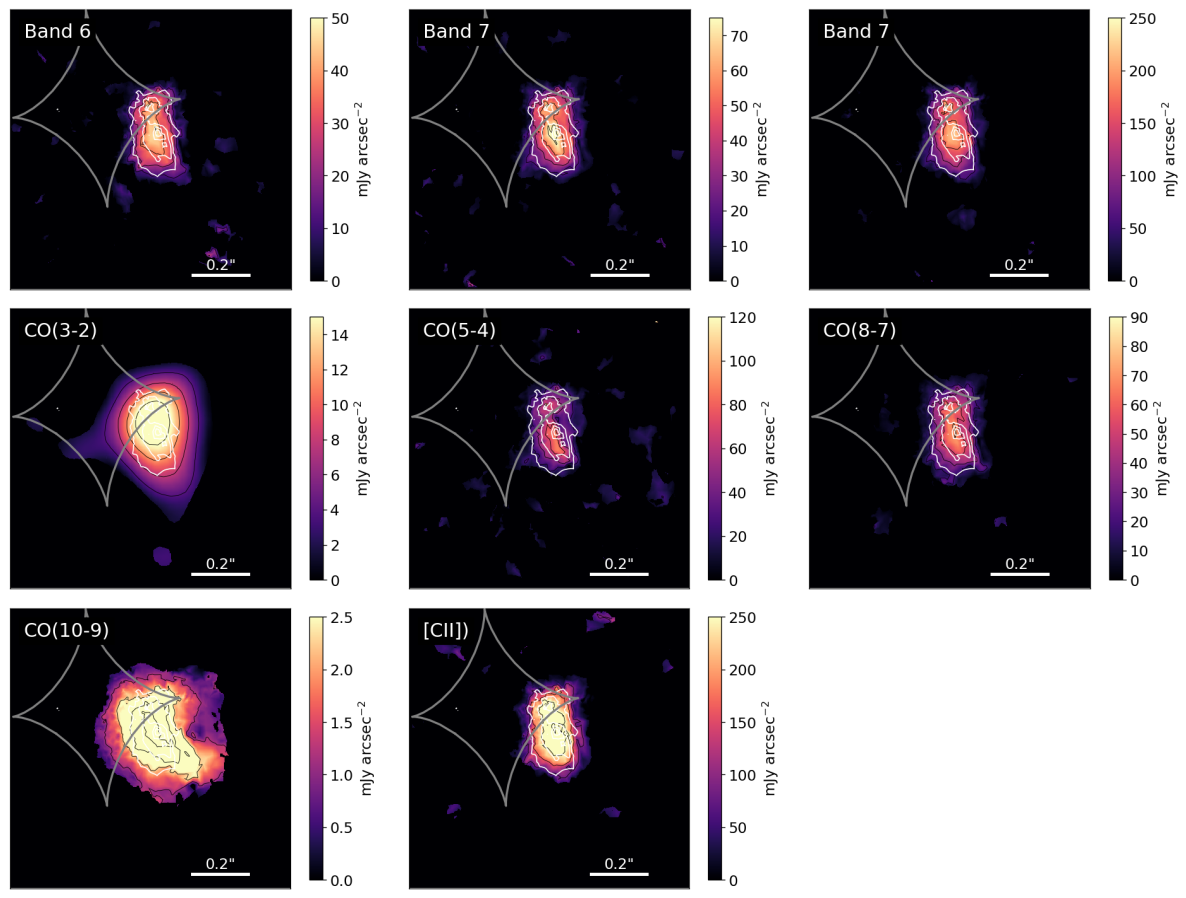}

\caption{\textit{Upper:} As Fig.~\ref{fig:app_B1}, but for an input source matching the FIR surface brightness distribution from Fig.~\ref{fig:sigma_SFR}. The white contours trace the 20, 40, 60 and 80~per cent of the input source peak surface brightness. \label{fig:app_B2}}
\end{centering}
\end{figure*}

We consider two different source-plane surface brightness distributions: (1) a uniform surface-brightness elliptical source with major/minor axis of 3.4$\times$2.4~kpc (corresponding to the FIR-bright source size at 45~deg inclination) and (2) a surface-brightness distribution matching the FIR luminosity map. These sources are then forward-lensed using the lens model for SDP.81. We then create mock ALMA observations for each tracer by setting the sky-plane flux density to the observed value (see Table~\ref{tab:image-plane}), using the same $uv$-plane coverage and adding the observed noise-per-baseline. Finally, we perform the full lens-modelling procedure for the each dataset. The reconstructed sources are shown in Figures~\ref{fig:app_B1} (uniform) and \ref{fig:app_B2} (FIR-like). 

First, we estimate the systematic error on the reconstructed source-plane surface brightness distribution, which we use in Section~\ref{sec:pdr_resolved}. Namely, we compare the input surface brightness for the uniform-brightness model (Fig.~\ref{fig:app_B1}) to the mean inferred surface brightness, integrated over the extent of the input source. The input and inferred surface brightness values, and the corresponding uncertainites are listed in Table~\ref{tab:app_b}.

For the continuum data, [\ion{C}{ii}] and CO(5--4) and (8--7) line, the resulting uncertainty is $\sim$10~per cent, i.e. comparable to the flux calibration uncertainty. However, for the CO(3--2) and (10--9) lines, which are observed at lower S/N, the fractional uncertainty is 65 and 30~per cent, respectively.

Second, we consider the reliability of the inferred source morphology:

\begin{itemize}

\item FIR continuum: the reconstructed Bands 6, 7 and 8 data recover the input source structure reliably, including the positions and separation of FIR-bright clumps in Fig.~\ref{fig:app_B2}.

\item CO(3--2): the low S/N and curvature regularization cause the source to be overly extended in the east-west (i.e.,\,radial) direction; the magnification in this direction is $\sim1$. The mean surface brightness is recovered within 20~per cent. However, neither of our mock reconstructions shows the increase in the CO(3--2) luminosity towards the north of the source, seen in the real data (Fig.~\ref{fig:reconstructions}). We therefore conclude that the observed variations in the CO(3--2) luminosity (and $\Sigma_\mathrm{H_2}$ in the depletion time analysis) is physical.

\item CO(5--4) and (8--7): both the uniform and FIR-like source morphologies are reliably recovered. The surface brightness bias is somewhat larger for the CO(8--7) line due to its lower S/N (15 per cent, compared to 10~per cent for CO(5--4). This confirms the robustness of the source-plane structures from \citet{Rybak2015b}, namely the extent of the CO(5--4)/(8--7) emission, and their offsets with respect to each other, and the FIR continuum (Fig.~\ref{fig:reconstructions}).

\item CO(10--9): the reconstructed source shows spurious structure for both mock source realizations. However, the observed concentration of the CO(10--9) emission to the north of the source is not (spuriously) reproduced in either of the mock source reconstructions; we thus consider it to be a physical. On the other hand, the low-surface brightness CO(10--9) to the south-west in Fig.~\ref{fig:reconstructions} is reminiscent of the spurious structures seen in the mock reconstructions. As this faint component contributes negligibly to the CO SLED and PDR modelling results in this paper, we do not exclude it from our analysis. However, we note that given the S/N of the data at hand, this could be a modelling artifact.

\item [\ion{C}{ii}]: both the uniform and FIR-like source are reliably recovered. In particular, for the FIR-based source, the positions of the FIR surface brightness maxima are recovered in the reconstructed maps. This confirms that the decline in [\ion{C}{ii}] surface brightness in these regions seen in the real data is physical, rather than a modelling artifact.

\end{itemize}

\begin{table}
\caption{Systematic error on the mean source surface brightness $\delta$ estimated from the mock source reconstructions in Fig.~\ref{fig:app_B1}. The columns list the input (uniform) surface brightness ($S_\mathrm{in}$), the measured mean surface brightness with 1$\sigma$ uncertainties ($\langle S_\mathrm{out} \rangle$), and the fractional systematic error $\delta$. \label{tab:app_b}}
\begin{center}
 \begin{tabular}{@{}l|ccc@{}}
 \hline \hline
Tracer & $S_\mathrm{in}$ & $\langle S_\mathrm{out} \rangle $ & $\delta$ \\
 & [mJy arcsec$^{-2}$] & [mJy arcsec$^{-2}$] & [per cent] \\
 \hline
CO(3-2) & 9.0 & 14.8$\pm$2.8 & 65\\
CO(5-4) & 24 & 22$\pm$7 & 10 \\
CO(8-7) & 24 & 20$\pm$4 & 20\\
CO(10-9) & 2.5 & 1.9$\pm$0.5 & 30\\
{[}\ion{C}{ii}{]} & 91 & 97$\pm$20 & 7 \\
Band~6 cont. & 13.1 & 12.9$\pm$3.8 & 2 \\
Band~7 cont. & 19 & 20$\pm$6 & 5 \\
Band~8 cont. & 55 & 61$\pm$14 & 11 \\
FIR & -- & -- & 7 \\
 \hline
 \end{tabular}
\end{center}
\end{table}

\section{Sensitivity of G, n to inferred line ratios}
\label{sec:appendix_C}

\begin{figure}
\begin{centering}
\includegraphics[width=8cm, clip=true]{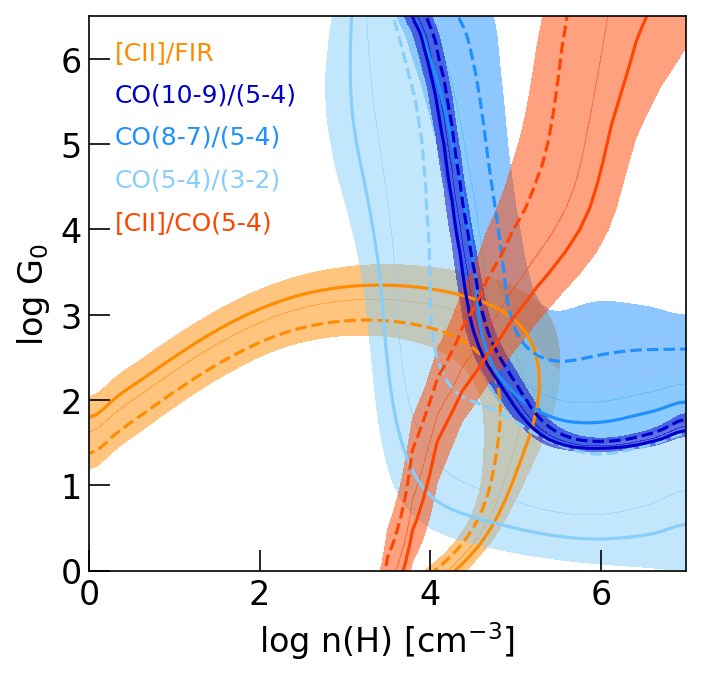}

\caption{Sensitivity of $G$ and $n$(H) values to inferred line ratios. The isocontours correspond to typical line ratios measured in SDP.81 (solid lines), and increased by a factor of 2 (dashed lines). The shaded regions correspond to an assumed 30~per cent uncertainty on line ratios. \label{fig:pdr_models_sensitivity}}
\end{centering}
\end{figure}

We now assess the robustness of $G$ and $n$ inferred from the PDR modelling to the bias in the reconstructed source-plane brightness distribution. In other words, if the lens modelling introduces a bias in the reconstructed source, how much will this affect the inferred $G$ and $n$? Fig.~\ref{fig:pdr_models_sensitivity} shows the $G$-$n$ space isocontours corresponding to 1$\times$ and 2$\times$ of the mean line ratios in SDP.81, using the {\sc PDRToolbox} models \citep{Kaufman1999}. As the systematic error the inferred line ratios is less than unity (Appendix~\ref{sec:appendix_B}), a change by a factor of two thus overestimates the actual uncertainty. As the line ratios considered are either between different species or non-neighbouring CO rotational transitions, a change by a factor of two results in relatively minor ($\leq$0.5~dex) shifts in the $G$/$n$(H) space. The only exception is the CO(5--4)/(3--2) ratio; however, given the relatively large associated uncertainty (Fig.~\ref{fig:pdr_models_global}), we do not expect the inferred PDR properties to be significantly biased.

\bibliographystyle{mnras}

\bibliography{references}


\bsp	
\label{lastpage}
\end{document}